\documentclass{aa}
\usepackage{graphicx}
\begin{document}
   \title{Mid-Infrared line diagnostics of Active Galaxies
          }

   \subtitle{A spectroscopic AGN survey with ISO-SWS
   \thanks{Based on observations with ISO, an ESA project with
                  instruments funded by ESA member states (especially
                  the PI countries: France, Germany, the Netherlands and
                  the United Kingdom) and with participation of ISAS and
                  NASA.} }

   \author{E. Sturm \inst{1}
          \and{D. Lutz \inst{1}}
          \and{A. Verma \inst{1}}
          \and{H. Netzer \inst{2}}
          \and{A. Sternberg \inst{2}}
          \and{A.F.M. Moorwood \inst{3}}
          \and{E. Oliva \inst{4}}
          \and{R. Genzel \inst{1}}
		   }

   \offprints{sturm@mpe.mpg.de}

   \institute{Max-Planck-Institut f\"ur extraterrestrische Physik,
                 Postfach 1312, D-85741 Garching, Germany
               \and School of Physics and Astronomy and Wise Observatory, Raymond and Beverly 
                 Sackler Faculty of Exact Sciences, Tel Aviv University, Ramat Aviv, Tel Aviv 69978, Israel 
               \and European Southern Observatory, Karl-Schwarzschild-Strasse 2, 85748 Garching, Germany
               \and Osservatorio Astrofisico di Arcetri, Largo E. Fermi 5, I-50125 Firenze, Italy, 
                 and Telescopio Nazionale Galileo \& Centro Galileo Galilei, calle Alvarez de Abreu 70, E-38700                  		     Santa Cruz de La Palma, TF Spain
              }

   \date{Received  / accepted }

   \abstract{
We present 
medium resolution (R$\sim$1500) 
ISO-SWS 2.4--45$\mu$m spectra of a sample of 29 galaxies with active nuclei. This data set
is rich in fine structure emission lines tracing the narrow line regions and \mbox{(circum-)}nuclear star 
formation regions, and it 
provides a coherent spectroscopic reference for future extragalactic studies in the mid-infrared.
We use the data set to briefly discuss the physical conditions in the narrow line regions 
(density, temperature, excitation, line profiles) and to test for possible differences
between AGN sub-types. 
Our main focus is on   
new tools for determining the properties of dusty galaxies and on the AGN-starburst 
connection.
We present mid-IR line ratio diagrams which can be used
to identify composite (starburst + AGN) sources and to
distinguish between emission excited by active nuclei and emission from
(circum-nuclear) star forming regions. 
For instance, line ratios of high to low excitation lines like
[O\,IV]25.9$\mu$m/[Ne\,II]12.8$\mu$m, that have been used to probe for AGNs in dusty objects,
can be examined in more detail and with better statistics now.
In addition, we present two-dimensional diagnostic diagrams that are fully analogous to 
classical optical diagnostic diagrams, but better suited for objects with high extinction.
Finally, we discuss correlations of mid-infrared line fluxes to
the mid- and far-infrared continuum. We compare these relations to similar relations in starburst
galaxies in order to examine the contribution of AGNs to the bolometric luminosities of their
host galaxies. The spectra are available in electronic form from the authors.

}  

  \maketitle


\section{Introduction}
 \label{s:introduction}

Mid-infrared spectra contain a
large number of atomic, ionic and molecular lines  along with various 
solid-state and dust features in grains of different sizes.
They are therefore an excellent tool to study the nature of infrared
bright galaxies, such as galaxies with active galactic nuclei (AGNs).
For instance, mid-infrared (MIR) continua can be compared to 
torus models (e.g. D. Alexander et al. 1999, Siebenmorgen et al. 1997), whereas
ice absorption features allow the study of the composition of the ISM and
the extinction law in AGNs (Chiar et al. 2000, Spoon et al. 2002).
Clavel et al. (2000) have used aromatic infrared bands (AIBs, often identified as `PAHs' - Polycyclic
Aromatic Hydrocarbons) 
as a new tool for testing unified schemes in Seyfert 1 and Seyfert 2 galaxies,
and Lutz et al. (2000a) have searched for broad line components of hidden broad line regions (BLRs).
For a more complete, recent review see Genzel \& Cesarsky (2000). 
A good overview of the global characteristics of MIR AGN spectra (like shapes, features, 
relative line intensities) can be obtained from 
the complete 2.5 to 45 $\mu$m spectra
of several AGN and starburst prototypes presented in Sturm et al. (2000), in combination
with the detailed and comprehensive line spectra of NGC 1068, NGC 4151, and the Circinus 
galaxy published in Lutz et al. (2000b), Sturm et al. (1999) and Moorwood et al. (1996). 

\begin{table*}
\caption[]{\label{tab:targets} The sample}
\begin{flushleft}
\begin{tabular}{llllllllll}
\hline\noalign{\smallskip}
\rule[-2mm]{0mm}{2mm}Nr. & Galaxy &   Type         &    z   & S$_{12}^\dagger$ & S$_{25}^\dagger$ & 
S$_{60}^\dagger$ & S$_{100}^\dagger$ &  D$^{\dagger\dagger}$ & L$_{IR}^{\dagger\dagger\dagger}$ \\ \hline     
1  & Cen A                        &   Sy2 + SB     &    0.001828  & 13.3 & 17.3 & 162 & 314 &   4 &  0.9  \\
2  & Circinus                     &   Sy2 + SB     &    0.001498  & 18.9 & 68.4 & 248 & 314 &   3 &  0.7  \\ 
3  & M 51                         &   Sy2 + SB     &    0.001544  & 11.0 & 17.5 & 109 & 292 &   6 &  1.3  \\ 
4  & M 87                         &   Sy           &    0.004360  &  0.2 &  0.2 & 0.4 & 1.0 &  17 &  0.1  \\ 
\rule[-3mm]{0mm}{3mm}5  & Mkn 1 
                     (=NGC 449)   &   Sy2          &    0.015944  &  1.9 &  0.9 & 2.5 & 2.9 &  65 &  9.5  \\ 
6  & Mkn 3                        &   Sy2          &    0.013509  &  0.7 &  2.9 & 3.8 & 3.4 &  54 &  6.1  \\ 
7  & Mkn 6                        &   Sy1.5        &    0.018813  &  0.2 &  0.7 & 1.2 &$<$1.7& 76 &  3.6  \\ 
8  & Mkn 335                      &   Sy1.2        &    0.025785  &  0.4 &  0.3 & 0.3 & 0.5 & 104 &  4.4  \\ 
9  & Mkn 463                      &   (2x) Sy2     &    0.050802  &  0.5 &  1.6 & 2.2 & 1.9 & 202 & 51.0  \\ 
\rule[-3mm]{0mm}{3mm}10 & Mkn 509 &   Sy1.2        &    0.034397  &  0.3 &  0.7 & 1.4 & 1.5 & 139 & 14.0  \\ 
11 & Mkn 573                      &   Sy2          &    0.017259  &  0.3 &  0.9 & 1.2 & 1.4 &  69 &  3.5  \\
12 & Mkn 938 (=NGC 34)            &   Sy2          &    0.019784  &  0.4 &  2.4 &16.6 &17.2 &  80 & 28.0  \\
13 & Mkn 1014                     &   Sy1          &    0.163099  &  0.1 &  0.5 & 2.2 & 2.2 & 677 &250.0  \\ 
14 & NGC 613                      &   Sy/Radio + SB&    0.004920  &  1.0 &  2.5 &  22 &  50 &  20 &  3.0  \\
\rule[-3mm]{0mm}{3mm}15 & NGC 1068&   Sy2          &    0.003829  & 38.0 & 86.0 & 186 & 239 &  14 & 30.0  \\ 
16 & NGC 1275                     &   Sy2 + SB     &    0.017559  &  1.1 &  3.5 & 7.1 & 6.9 &  71 & 16.0  \\
17 & NGC 1365                     &   Sy1.8 + SB   &    0.005457  &  3.3 &  1.1 &  76 & 142 &  22 & 12.0  \\
18 & NGC 3783                     &   Sy1          &    0.009730  &  0.8 &  2.5 & 3.3 & 4.9 &  34 &  2.5  \\
19 & NGC 4051                     &   Sy1.5        &    0.002418  &  0.9 &  1.6 & 7.1 &23.9 &  10 &  0.4  \\
\rule[-3mm]{0mm}{3mm}20 & NGC 4151&   Sy1.5        &    0.003319  &  1.8 &  4.7 & 6.9 &10.2 &  20 &  1.7  \\ 
21 & NGC 5506                     &   Sy1.9/NLXG   &    0.006181  &  1.3 &  3.6 & 8.4 & 8.9 &  23 &  2.0  \\  
22 & NGC 5643                     &   Sy2          &    0.003999  &  1.1 &  3.7 &19.5 &38.2 &  16 &  1.8  \\
23 & NGC 7469                     &   Sy1.2 + SB   &    0.016398  &  1.3 &  5.8 &25.9 &34.9 &  65 & 29.0  \\ 
24 & NGC 7582                     &   Sy2/NLXG + SB&    0.005254  &  1.4 &  6.3 &  48 &72.8 &  20 &  5.0  \\ 
\rule[-3mm]{0mm}{3mm}25 & PKS 2048-57&   Sy2       &    0.011348  &  1.1 &  3.9 & 4.3 & 4.2 &  45 &  6.1  \\ 
26 & Tol 0109-383 (=NGC 424)      &   Sy1.8        &    0.011661  &  1.1 &  1.7 & 1.8 & 1.8 &  47 &  3.7  \\ 
27 & I Zw 1                       &   Sy1          &    0.061086  &  0.5 &  1.2 & 2.2 & 2.6 & 248 & 71.0  \\
28 & I Zw 92 (=Mkn 477)           &   Sy2 + SB     &    0.037799  &  0.1 &  0.5 & 1.3 & 1.9 & 153 & 12.0  \\
29 & 3C120 (=II Zw 14)            &   Sy1          &    0.033010  &  0.3 &  0.6 & 1.3 & 2.8 & 135 & 13.5  \\
\noalign{\smallskip}
\hline
\end{tabular}
\end{flushleft}
$\dagger$ = IRAS (FSC) fluxes at 12, 25, 60 and 100 $\mu$m in Jy\\
$\dagger\dagger$ = distance in Mpc, for H$_0$=75km/s/Mpc and q$_0$=0.5\\
$\dagger\dagger\dagger$ = 8-1000$\mu$m luminosity in 10$^{10}$[L$_\odot$], L$_{\rm IR}$=
5.6xD$_{\rm Mpc}^2$(S$_{100}$+2.58S$_{60}$+5.16S$_{25}$+13.48S$_{12}$), see Sanders \&Mirabel (1996)
\end{table*}

\begin{table*}
\caption[]{\label{tab:fluxes-fine} Observed fine structure line fluxes and upper 
           limits (in 10$^{-20}$W/cm$^2$).}
{\footnotesize
\begin{flushleft}
\begin{tabular}{lllllllll}
\hline\noalign{\smallskip}
\rule[-2mm]{0mm}{2mm}Galaxy        &        
                     [Si\,IX] & [Mg\,VIII] &     [Si\,IX] & [Mg\,IV] &  [Fe\,II] & [Mg\,VII] & [Mg\,V] &
                     [Ar\,II]  \\ \hline     
$\lambda_{\rm rest,vac}$($\mu$m) &  
                       2.584 & 3.028     &           3.936   & 4.487   & 5.340   & 5.503    & 5.610  &                                                                      			     6.985    \\ 
E$_{\rm ion}$ (eV)$^1$&  303.2   & 224.9 &           303.2   & 80.1    & 7.9     & 186.5    & 109.2  &                             		          15.8   \\
Resolution$^4$    &  100     & 120       &           150     & 130     & 250     & 230      & 230    &
                        190   \\ 
\rule[-3mm]{0mm}{3mm}Aperture$^2$&
                      14x20  & 14x20     &           14x20   & 14x20   & 14x20   & 14x20    & 14x20  & 
                      14x20  \\
Cen A       &         ...    & $<$0.5    &           $<$0.4  & ...     & 1.6     & 1.5      & 0.9$^3$  &                         
                      4.9    \\
Circinus    &         1.7    & 6.2       &           4.9     & 1.7     & 8.3     & 8.0      & 6.7    &                                                                
			   27.6    \\
M51         &         ...    & $<$0.3    &           ...     & ...     & ...     & ...      & $<$1.2 &                       
			    ...    \\  
M87         &         ...    & ...       &           ...     & ...     & ...     & ...      & $<$1.0 & 
                      ...    \\ 
\rule[-3mm]{0mm}{3mm}Mkn1& 
                      ...    & ...       &           ...     & ...     & ...     & ...      & $<$2.0 & 
                      ...    \\ 
Mkn3        &         ...    &  ...      &           ...     & ...     & ...     & ...     & ...    &
                      ...    \\
Mkn6        &       $<$0.6   &$<$0.5     &           ...     &$<$1.5   & ...     & ...     & ...    &
                      ...     \\
Mkn335      &         $<$0.5 & $<$0.3    &           $<$0.2  & ...     & ...     & $<$1.8  & ...    &                                             
                      ...    \\ 
Mkn463      &         ...    & ...       &           ...     & ...     & ...     & $<$0.8   & $<$0.9 &                       
                      ...    \\ 
\rule[-3mm]{0mm}{3mm}Mkn509 & 
                      ...    &  0.4      &           ...     & ...     & ...     & ...     & ...    &
                      ...    \\ 
Mkn573      &        $<$0.8  &  0.6      &           ...     & ...     & ...     & ...     & $<$1.0 &
                      ...    \\  
Mkn938      &        $<$1.4  &  ...      &           ...     & ...     & ...     & ...     & ...    &
                      ...    \\ 
Mkn1014     &         ...    & $<$1.3    &           ...     & ...     & ...     & ...      & $<$1.5 & 
                      ...    \\ 
NGC613      &         ...    &  ...      &           ...     & ...     & ...     & ...     & ...    &
                      ...    \\
\rule[-3mm]{0mm}{3mm}NGC1068 &  
                      3.0    & 11.0      &           5.4     & 7.4     & 5.0     &13.0      &18.0    &                      
                     13.0    \\  
NGC1275     &         $<$0.7 & ...       &           $<$0.2  & ...     & $<$1.0  & ...      & $<$1.3 & 
                      0.6    \\
NGC1365     &         $<$0.5 & $<$0.3    &           $<$0.3  & $<$1.5  & ...     & $<$1.8   & $<$1.5 &                      
                     14.1    \\
NGC3783     &         ...    & ...       &           ...     & ...     & ...     & ...      & $<$1.0 & 
                      ...    \\
NGC4051     &         ...    &  ...      &           ...     & ...     & ...     & ...     & ...    &
                      ...    \\
\rule[-3mm]{0mm}{3mm}NGC4151&
                      0.2    & 0.6       &           0.4     & 0.3     & ...     & $<$1.0  & $<$1.5 & 
                      ...    \\
NGC5506     &         $<$1.4 & $<$0.4    &           $<$0.3  & $<$1.8  & $<$2.0  & $<$2.4  & $<$1.8 & 
                      3.2    \\
NGC5643     &         ...    &  ...      &           ...     & ...     & ...     & ...     & $<$2.0 &                                                                     
                      ...    \\
NGC7469     &         $<$0.8 & $<$0.2    &           $<$0.4  & $<$1.0  & ...     & $<$2.3  & $<$2.5 & 
                      6.7    \\
NGC7582     &         $<$0.5 & $<$0.3    &           $<$0.3  & $<$1.2  & $<$1.5  & $<$0.9  & $<$1.5 & 
                      ...    \\
\rule[-3mm]{0mm}{3mm}PKS2048-57  &         
                      ...    & 0.4       &           0.2     & ...     & ...     & $<$1.4  & $<$1.4 &                       
                      ...    \\
TOL0109-383 &         ...    & $<$0.2    &           $<$0.2  & ...     & ...     & $<$1.1   & ...   & 
                      ...    \\ 
I Zw 1      &         ...    & ...       &           ...     & ...     & ...     & ...      & $<$0.6&                                                	                ...    \\
I Zw 92     &        $<$0.4  &  ...      &           ...     &$<$1.5   & ...     & ...     & ...    &
                      ...    \\ 
3C120       &        $<$0.4  &$<$0.4     &           ...     &$<$0.8   & ...     & ...     & ...    &
                      ...    \\ 
\noalign{\smallskip}
\hline
\end{tabular}
\end{flushleft}
1 = lower ionization potential of the stage leading to the transition\\
2 = in arcsec\\
3 = uncertain detection, either because of noise ($\approx 3\sigma$) or offset in wavelength
larger than 100 km/s compared to similar lines in the same object\\
4 = in km/s
}
\end{table*}

\begin{table*}
\addtocounter{table}{-1}
\caption[]{\it continued}
{\footnotesize
\begin{flushleft}
\begin{tabular}{lllllllllll}
\hline\noalign{\smallskip}
\rule[-2mm]{0mm}{2mm}Galaxy        &        
                       [Ne\,VI]&[Fe\,VII]& [Ar\,III]/[Mg\,VII]&[Fe\,VII]&[S\,IV]& [Ne\,II]&[Ne\,V] & [Ne\,III] & 
                       [P\,III]&[Fe\,II] \\ \hline     
$\lambda_{\rm rest,vac}$($\mu$m) &  
                       7.652   & 7.815   & 8.991            & 9.527  &10.551   & 12.814 & 14.322& 15.555   & 
                       17.885  & 17.936  \\ 
E$_{\rm ion}$ (eV)$^1$&  126.2 & 99.1    & 27.6/186.5       & 99.1   & 34.8    & 21.6   &97.1   & 41.0     & 
                       19.8    & 7.9     \\
Resolution$^4$    &    165     & 160     & 150              & 140    & 115     & 190    & 160   & 150      &
                       125     & 125     \\ 
\rule[-3mm]{0mm}{3mm}Aperture$^2$&
                      14x20    &  14x20  &    14x20         & 14x20  &14x20    & 14x27  & 14x27 &  14x27   &
                       14x27   &  14x27  \\
Cen A       &         2.0      & ...     & 0.9              & ...    & 1.4     & 22.1   &  2.7  & 14.1     &  
                       0.4     &   0.2   \\
Circinus    &        36.3      & 1.5     & 8.3              & 1.1    &12.7     & 90.0   & 31.7  & 33.5     &
                        1.3$^3$  & $<$1.0  \\
M51         &         $<$0.3   & ...     & $<$0.5           & ...    &$<$0.8   &  7.0   & $<$0.2&  2.9     & 
                        ...    &   ...   \\  
M87         &           ...    & ...     & $<$0.4           & ...    &...      &  ...   &  ...  & ...      &  
                        ...    &   ...   \\ 
\rule[-3mm]{0mm}{3mm}Mkn1& 
                        ...    & ...     & $<$0.9           & ...    &...      &  ...   &  ...  & ...      & 
                        ...    &   ...   \\ 
Mkn3        &           4.5    &  ...    & ...              & ...    & ...     & 4.7    & 4.6   &12.3      & 
                        ...    &  ...    \\
Mkn6        &         $<$0.5   &  ...    & ...              & ...    & ...     & 0.8    &$<$0.6 & 1.8$^3$    & 
                        ...    &  ...    \\
Mkn335      &         $<$0.3   & ...     & ...              & ...    &...      & $<$0.7 & $<$0.5& $<$0.8   &  
                        ...    &   ...   \\ 
Mkn463      &           ...    & ...     & 0.3              & ...    &...      &  1.3   & 1.4   & ...      &  
                        ...    &   ...   \\ 
\rule[-3mm]{0mm}{3mm}Mkn509  &       
                        0.4    &  ...    & ...              & ...    & ...     & 2.0    & ...   & ...      & 
                        ...    &  ...    \\ 
Mkn573      &           1.2    &  ...    & ...              & ...    & ...     &$<$1.3  & 1.8   & 2.4      & 
                        ...    &  ...    \\  
Mkn938      &           ...    &  ...    & ...              & ...    & ...     & 5.7    &$<$0.4 &$<$1.0    & 
                        ...    &  ...    \\ 
Mkn1014     &           ...    & ...     & ...              & ...    &...      &  0.4$^3$ &$<$0.4 & ...      &  
                        ...    &   ...   \\ 
NGC613      &          $<$0.3  &  ...    & ...              & ...    & ...     & 4.0    &$<$0.8 &$<$6.8    & 
                        ...    &  ...    \\
\rule[-3mm]{0mm}{3mm}NGC1068 & 
                      110.0    & 3.0     &25.0              & 4.0   &58.0      & 70.0   & 97.0  &160.0     &  
                        ...    & $<$10.0 \\  
NGC1275     &         $<$0.5   & ...     & $<$0.5           & ...   &$<$0.3    &  2.9   &$<$0.7 & 1.0      &  
                        0.4$^3$  &$<$0.7   \\
NGC1365     &           2.5    & ...     & 1.1              & ...   &2.6       & 40.9   & 2.5   & 7.7      &  
                        ...    &   ...   \\
NGC3783     &           ...    & ...     & 0.5$^3$            & ...   &...       &  ...   & 2.1   & ...      &  
                        ...    &   ...   \\
NGC4051     &           ...    &  ...    & ...              & ...   & ...      & 2.5    &$<$1.0 & 3.0      & 
                        ...    &  ...    \\
\rule[-3mm]{0mm}{3mm}NGC4151 &
                        7.8    & ...     & 2.2              & ...  &11.3       & 11.8   & 5.5   & 20.7     &  
                        ...    &   ...   \\
NGC5506     &           2.2    & ...     & 0.7$^3$            & ...  &5.4        &  5.9   & 2.6   & 5.8      &  
                        ...    &   ...   \\
NGC5643     &           ...    & ...     & 0.9              & ...  &...        & ...    & ...   & ...      &  
                        ...    &  ...    \\
NGC7469     &           1.1    & ...     & 0.5$^3$            & ...  & 0.9       & 22.6   & $<$1.5& 2.2      &  
                        ...    & ...     \\
NGC7582     &           2.3    & ...     & 1.0              & ...  &1.8        & 14.8   &2.2    & 6.7      &  
                        ...    & $<$0.6  \\
\rule[-3mm]{0mm}{3mm}PKS2048-57  &      
                        2.8    & ...     & 1.0              & ...  &...        &  2.1$^3$ &3.6    & 5.9      &  
                        ...    & ...     \\
TOL0109-383 &           1.2    & ...     & ...              & ...  &..         & $<$1.2 & 0.8   & $<$1.6   &  
                        ...    &  ...    \\ 
I Zw 1      &           $<$0.3 & ...     & ...              & ...  &...        & 0.65   & 0.27  & ...      & 
                        ...    &  ...    \\
I Zw 92     &           $<$0.7 &  ...    & ...              & ...    & ...     & 2.4    &$<$1.5 & 1.6      & 
                        ...    &  ...    \\ 
3C120       &           1.7    &  ...    & ...              & ...    & ...     & 0.9    & 1.9   & 2.0      & 
                        ...    &  ...    \\
\noalign{\smallskip}
\hline
\end{tabular}
\end{flushleft}
}
\end{table*}

\begin{table*}
\addtocounter{table}{-1}
\caption[]{\it continued}
{\footnotesize
\begin{flushleft}
\begin{tabular}{llllllllllll}
\hline\noalign{\smallskip}
\rule[-2mm]{0mm}{2mm}Galaxy        &        
                     [S\,III]&[Fe\,III] &[Fe\,I] &[Ne\,V]   & [S\,I] &[O\,IV]& [Fe\,II] & [S\,III] & 
                     [Si\,II]& [Fe\,II] &[Ne\,III] \\ \hline   
$\lambda_{\rm rest,vac}$($\mu$m) &
                     18.713  & 22.925   & 24.042 & 24.318   & 25.249 &25.890 & 25.988   & 33.418  & 
                     34.815  & 35.349   & 36.014   \\ 
E$_{\rm ion}$ (eV)$^1$&  23.3& 16.2     & 0.0    & 97.1     & 0.0    &54.9   &  7.9     & 23.3    &  
                      8.2    &  7.9     & 41.0     \\ 
Resolution$^4$    &  120     & 250      &  230   & 230      & 225    & 215   &  215     &  195    &
                     185     &  185     &  185     \\  
\rule[-2mm]{0mm}{3mm}Aperture$^2$&  
                     14x27   & 14x27    & 14x27  & 14x27    & 14x27  & 14x27 &  14x27   & 20x33   &  
                     20x33   &  20x33   &  20x33   \\ 
Cen A       &         6.4    &  0.3     & $<$0.2 & 2.0      & $<$0.3 &  9.8  &  1.2     & 22.3    &  
                      54.5   &  ...     &   1.4    \\ 
Circinus    &        35.2    &  1.2     & $<$1.0 & 21.8     & $<$0.5 & 67.9  &  5.9     & 93.2    & 
                     151.0   &  1.5     &   4.0    \\ 
M51         &         1.0    &  ...     &  ...   & ...      & ...    &  1.9  &  1.0     &  4.6    &   
                      9.6    &  ...     &  ...     \\  
M87         &        $<$0.6  &  ...     &  ...   & ...      &  ...   & $<$0.2&  0.5     & $<$1.4  & 
                     $<$1.2  &  ...     &  ...     \\ 
\rule[-3mm]{0mm}{3mm}Mkn1&
                     $<$0.8  &  ...     &  ...   & ...      &  ...   &  2.8  &$<$0.4    & $<$1.0  & 
                     $<$1.3  &  ...     &  ...     \\ 
Mkn3        &         ...    &  ...     & ...    & 3.4      & ...    & 12.6  & ...      & 3.0     & 
                      ...    &  ...     & $<$2.5   \\
Mkn6        &         ...    &  ...     & ...    &$<$0.5    & ...    & 1.6$^3$ & ...      &$<$1.8   & 
                      ...    &  ...     &  ...     \\
Mkn335      &         ...    &  ...     &  ...   &$<$0.3    &  ...   &  1.3  &$<$0.3    &  ...    &   
                      ...    &  ...     & $<$0.5   \\ 
Mkn463      &        $<$0.8  &  ...     &  ...   & ...      &  ...   &  4.3  & ...      &  1.2    &   
                       2.0   &  ...     &  ...     \\ 
\rule[-3mm]{0mm}{3mm}Mkn509  & 
                     $<$1.5  &  ...     & ...    &$<$0.6    & ...    & 1.8   & ...      & ...     & 
                      ...    &  ...     &$<$2.0    \\ 
Mkn573      &         ...    &  ...     & ...    & ...      & ...    & 7.9   & ...      & 2.9     & 
                      ...    &  ...     &  ...     \\  
Mkn938      &         ...    &  ...     & ...    & ...      & ...    & ...   & ...      & ...     & 
                      ...    &  ...     &  ...     \\ 
Mkn1014     &        $<$0.6  &  ...     &  ...   & ...      &  ...   & 2.4$^3$ &$<$1.0    & $<$1.6  & 
                     $<$1.3  &  ...     &  ...     \\ 
NGC613      &         ...    &  ...     & ...    &$<$0.7    & ...    & ...   & ...      & ...     & 
                      ...    &  ...     & $<$1.0   \\
\rule[-3mm]{0mm}{3mm}NGC1068 &  
                      40.0   &  ...     &  ...   &70.0      &  ...   &190.0  &8.0$^3$     & 55.0    &  
                      91.0   & $<$1.4   &  18.0    \\  
NGC1275     &        $<$1.3  & $<$0.5   & $<$0.4 &$<$0.4    &  $<$0.5& $<$0.5&$<$0.4    & $<$1.7  &   
                       6.8   &  ...     & $<$1.0   \\
NGC1365     &         13.5   &  0.8     & $<$0.7 & 3.9      & $<$0.4 & 14.6  & 1.8      & 36.1    &  
                      73.8   &  ...     & $<$1.6   \\
NGC3783     &          2.0   &  ...     &  ...   &...       &  ...   &  3.8  &$<$0.5    & $<$2.0  &   
                       3.2   &  ...     &  ...     \\
NGC4051     &         ...    &  ...     & ...    & ...      & ...    & ...   & ...      & ...     & 
                      ...    &  ...     &  ...     \\
\rule[-3mm]{0mm}{3mm}NGC4151 &
                       5.4   & $<$0.2   & $<$0.4 &5.6       &  ...   & 20.3  &  0.4     &  8.1    &  
                      15.6   &  ...     &  3.5     \\
NGC5506     &          3.0   & $<$0.4   & $<$0.5 &2.4       &  ...   & 13.5  &0.6$^3$     &  8.1    &  
                      14.2   &  ...     &  1.2$^3$   \\
NGC5643     &          2.3   &  ...     & ...    &...       & ...    & 10.3  & ...      & 5.8     &  
                       5.5   &  ...     & ...      \\
NGC7469& 
                       9.2   &  0.9     & $<$0.4 &0.6$^3$     &  ...   &  3.4  & 0.6      & 10.4    &  
                      19.6   &  ...     & $<$1.0   \\
NGC7582     &          5.2   & $<$0.5   & $<$0.5 &3.1       &  ...   & 11.6  & 0.8      & 11.3    &  
                      21.8   & $<$1.4   &   1.0$^3$  \\
\rule[-3mm]{0mm}{3mm}PKS2048-57  &  
                       2.2   &  ...     &  ...   &2.7       &  ...    & 8.5  & 0.5      &  2.7    &   
                       3.5   &  ...     & $<$1.2   \\
TOL0109-383 &          ...   &  ...     &  ...   &0.5       &  ...   & 1.4   &$<$0.3    &  ...    &  
                       ...   &  ...     & $<$2.0   \\ 
I Zw 1      &        $<$0.5  &  ...     &  ...   &...       & ...    & $<$0.6& ...      & $<$1.0  &    
                     $<$1.8  &  ...     & $<$0.4   \\
I Zw 92     &         ...    &  ...     & ...    & ...      & ...    & 1.1   & ...      & 1.5     & 
                      ...    &  ...     & ...      \\
3C120       &         ...    &  ...     & ...    & ...      & ...    & 7.5   & $<$0.3   &$<$1.2   & 
                      ...    &  ...     &  ...     \\ 
\noalign{\smallskip}
\hline
\end{tabular}
\end{flushleft}
}
\end{table*}

%
\begin{table*}
\caption[]{\label{tab:fluxes-fine3} Additional fine structure line fluxes 
           and upper limits (in 10$^{-20}$ W/cm$^2$).}
{\footnotesize
\begin{flushleft}
\begin{tabular}{lllllllllllll}
\hline\noalign{\smallskip}
\rule[-2mm]{0mm}{2mm}Galaxy        &        
                   [Ca\,IV] & [Ca\,VII]& [Ca\,V]  & [Ar\,VI] & [Si\,VII]& [Na\,III]& [Na\,VI]&
                   [Ca\,V]  & [Mg\,V]  & [Fe\,II] & [Fe\,III] \\ \hline     
$\lambda_{\rm rest,vac}$($\mu$m) &  
                     3.207  &   4.087  & 4.158    &  4.530   &  6.492   &  7.318   &  8.611  & 
                     11.482 &  13.521  & 24.519   &   33.038  \\ 
Resolution$^4$     &  125   &   150    & 140      &   135    &  215     &   180    &  150    &
                        110 &  185     & 220      &    200    \\
E$_{\rm ion}$ (eV)$^1$& 50.9&   108.8  & 67.3     &  75.0    &  205.1   &  47.3    &  138.4  &
                      67.3  &  109.2   &    7.9   &   16.2    \\
\rule[-2mm]{0mm}{3mm}Aperture$^2$&
                     14x20  &  14x20   &  14x20   & 14x20    & 14x20    & 14x20    & 14x20   & 
                     14x20  &  14x27   & 14x27    & 20x30     \\
Circinus         &   $<$0.9 & $<$3.5   & $<$3.0   & 3.1      & $<$1.0   &  ...     &  ...    &
                     $<$0.8 & $<$1.5   & $<$0.5   & $<$5.0     \\
NGC1068          &    1.3   &  ...     & ...      & 15.0     &  ...     &  5.8     &$<$16.0  &
                      ...   &  ...     &  ...     &  ...       \\

\noalign{\smallskip}
\hline
\end{tabular}
\end{flushleft}
1 = lower ionization potential of the stage leading to the transition\\
2 = in arcsec\\
}
\end{table*}

In this paper we concentrate on narrow emission lines. The MIR fine structure emission lines cover 
a wide range of physical conditions, like excitation (0-300 eV), or 
critical density (10$^2$ $<$ n$_H$ $<$ 10$^7$ cm$^{-3}$). They are significantly less affected by 
extinction than UV, optical or near-IR lines, and 
their emissivities depend only weakly on electron temperature, in contrast to their optical and 
UV counterparts. Low excitation ionic fine 
structure lines (excitation potential $\le$ 50 eV) sample HII regions that are photoionized 
by massive stars (Spinoglio \& Malkan 1992, Voit 1992, Giveon et al. 2002). 
Some contribution from ionizing shocks is possible (e.g. Contini \& Viegas 2001). Ionic lines 
from species 
with excitation potentials up to $\approx$ 300 eV sample highly ionized or `coronal' gas and 
require the very hard radiation fields of AGNs, or fast ionizing shocks. 
Electron densities can be determined from ratios like [Ne\,V] 14/24 $\mu$m.
Line ratios from different ionization stages probe the hardness of the radiation field and 
the ionization parameter (e.g. [Ne\,II], [Ne\,III], [Ne\,V], [Ne\,VI]), but possible contributions
from nuclear starburst components have to be taken into account. Finally, 
hydrogen recombination lines and line ratios like [S III]33$\mu$m/18$\mu$m can be used to trace 
the MIR extinction.

The Infrared Space Observatory (ISO, Kessler et al. 1996) has  
provided access to this wealth of information. 
In the following we present a compilation of ISO-SWS 
emission line spectra and line fluxes of AGNs (Sect. \ref{s:results}). 
We use this data set to derive electron densities and temperatures
in the narrow line regions (NLRs) of AGNs (Sects. \ref{s:nlr_cond1} 
and \ref{s:nlr_cond2}),
and we compare the line ratios of highly ionized lines to the predictions of
photoionization models to constrain the spectral energy distribution (SED) of the 
central ionizing source and to test their consistency with AGN unification 
models (Sect. \ref{s:central_source}).
Furthermore, we study line profiles and their various degrees of asymmetry and differences
to the profiles of their optical counterparts (Sect. \ref{s:profiles}).  
In Sect. \ref{s:star_contrib} we develop a new diagram to
identify those sources in our sample where 
star forming regions within the SWS apertures contribute to the fluxes of 
low lying fine structure lines (`composite sources'). 
In order to establish tools for the distinction of AGNs and starbursts in 
distant dusty galaxies we construct mid-IR 
line ratio diagrams which are better suited for dusty objects than their
classical optical counterparts
(e.g. Baldwin et al. 1981; 
Veilleux \& Osterbrock 1987) 
in Sect. \ref{s:new_diag}.
Finally, we discuss correlations of mid-IR lines with far-IR  fluxes 
in the context
of the far-IR energy source and the starburst-AGN connection (Sect. \ref{s:fir_source}).
The rich spectrum of molecular hydrogen (H$_2$) in our data set is
discussed in a separate paper (Rigopoulou et al. 2002).

\section{Target Selection}
\label{s:targets}

Most of the objects discussed here have been observed as part of a guaranteed 
time ISO project on bright galactic nuclei (MPEXGAL), 
expanded by a solicited follow-up proposal (ZZCORLIN). 
They are well studied, nearby Seyfert galaxies, selected only on the basis
of their brightness and visibility to ISO. The sample is therefore not complete
in any sense.  
From the ISO archive we have added 9 targets which have been observed in other programmes, 
but for which fluxes have not been published 
to date. For three of these additional sources relative line strengths
have been previously published and studied by Spinoglio et al. (2000).
Table \ref{tab:targets} lists the names and 
classifications of all 29 galaxies of our sample. 
According to the classifications given in the NASA/IPAC Extragalactic Database 
(NED) the sample consists of 
13 Seyfert 1s (of various sub-types) and 14 Seyfert 2s. NGC 5506 and
NGC7582 are classified as NLXGs. No Seyfert type is given for M 87 and NGC 613.
For 8 additional Seyfert galaxies (from the CfA sample) SWS fluxes or flux limits 
of 4 coronal lines can be found in Prieto \& Viegas (2000).

Many active galaxies exhibit intense star formation in the (circum-)nuclear
region. In these cases some fraction of the measured fluxes of low lying fine structure
lines (excitation potential $\le$ 50 eV) will be produced by photoionization from stars 
rather than from the AGN. In Table \ref{tab:targets} we have indicated (Type=`Sy+SB') 
such composite sources
where star formation regions within the SWS apertures are important. 
This classification will be discussed in Sect. \ref{s:star_contrib}.



\section{Observation and Data Processing}
\label{s:obs}

The data have been obtained with the Short-Wavelength Spectrometer (SWS, de Graauw et al. 1996)
on board the Infrared Space Observatory (ISO), 
using the SWS02 and SWS06 observation modes. 
These modes provide line spectra with a resolving power of approximately 
1000--2500 in the range between 2.4 and 45 $\mu$m.
For Circinus and NGC 1068 a few lines have been added from SWS01 measurements
with reduced resolving power. The observations were centered on the nuclei.
Total observation times per source ranged from 10 minutes to 7 hours. 
Spectral lines at different wavelengths are observed with
three different aperture sizes, varying between 14\arcsec$\times$20\arcsec\/ and 
20\arcsec$\times$33\arcsec (see Table \ref{tab:fluxes-fine}).

We have processed the data using the SWS Interactive Analysis (IA) system
(Lahuis et al. 1998, Wieprecht et al. 1998) and the ISO Spectral Analysis
Package ISAP (Sturm et al. 1998) together with 
the associated calibration files available at the end of 2000/beginning of 2001. Dark current subtraction, scan direction 
matching, and flatfielding have been done interactively, and noisy 
detectors have been eliminated. Outliers were clipped, and the data of all 
12 detectors were averaged for each line, retaining the 
instrumental resolution. In a few cases some wavelength ranges were affected by fringes.
In these cases we have defringed their averaged spectra using the FFT or iterative 
sine fitting options of the defringe module within ISAP. 


\section{Results}
\label{s:results}

The spectra are displayed in Figs. \ref{F:figspec1} to \ref{F:figspec7}. 
The fine structure emission lines are generally resolved, having 
widths (FWHM) in the range between 200 and 600 km/s, i.e. line widths 
typical of narrow line regions of Seyfert galaxies. 
Measured line fluxes and upper limits are summarized in Table \ref{tab:fluxes-fine} 
(fine structure lines), Table \ref{tab:fluxes-fine3} (additional fine structure lines for few sources), 
and Table \ref{tab:fluxes-recomb} (hydrogen recombination lines). The spectra and
fluxes for NGC\,4151 and NGC \,1068 have already been published in Sturm et al. (1999) and Lutz et al. (2000b) respectively. We have repeated the fluxes (but not the spectra) 
here for convenience. Some of the
Circinus data have been published in Moorwood et al. (1996), but later on
more data have been taken. We have re-analyzed the old Circinus data with the newer calibration files in order to be consistent with the reduction of the additional observations.


\begin{table}
\caption[]{\label{tab:fluxes-recomb} Observed hydrogen recombination line fluxes (in 10$^{-20}$ W/cm$^2$).}
{\footnotesize
\begin{flushleft}
\begin{tabular}{llllll}
\hline\noalign{\smallskip}
\rule[-2mm]{0mm}{2mm}Galaxy        &        
                     Br\,$\beta$ & Pf\,$\gamma$& Br\,$\alpha$& Pf\,$\beta$& Pf\,$\alpha$ \\ \hline     
$\lambda_{\rm rest,vac}$($\mu$m) &  
                       2.625     &     3.741  &   4.052    &  4.654    &    7.460   \\ 
Resolution$^1$       &   145     &     155    &   150      &  135      &    180     \\
\rule[-2mm]{0mm}{3mm}Aperture$^2$&
                       14x20     &    14x20   &   14x20    &  14x20    &    14x20   \\
Cen A           &        0.9     &    ...     &     0.8    &   ...     &   0.2      \\                            
Circinus        &        3.2     &    ...     &     5.5    &   ...     &   2.1      \\
M51             &        $<$0.8  &    ...     &     ...    &   ...     &   ...      \\
Mkn335          &        $<$0.3  &    ...     &     ...    &   ...     &   ...      \\
\rule[-2mm]{0mm}{3mm}Mkn463 & 
                         $<$0.4  &    ...     &     ...    &   ...     &   ...      \\
Mkn1014         &        $<$0.3  &    ...     &     ...    &   ...     &   ...      \\
NGC1068         &        4.1     &    $<$0.4  &     6.9    &   $<$10.0 &   $<$3.0   \\
NGC1275         &        $<$0.2  &    ...     &     $<$2.2 &   ...     &   ...      \\
NGC1365         &        1.6     &    ...     &     2.8    &   ...     &   0.6      \\
\rule[-2mm]{0mm}{3mm}NGC3783 &
                         $<$0.8  &    ...     &     ...    &   ...     &   ...      \\
NGC4151         &        0.5     &    ...     &    0.9$^3$ &   ...     &   ...      \\ 
NGC5506         &      0.7$^4$   &    ...     &     1.2    &   ...     &   ...      \\ 
NGC5643         &      0.4$^5$   &    ...     &     ...    &   ...     &   ...      \\ 
NGC7469         &      0.75$^6$  &    ...     &     ...    &   ...     &   ...      \\
\rule[-2mm]{0mm}{3mm}NGC7582 &
                         0.9     &    ...     &     1.6    &   ...     &   ...      \\
PKS2048-57      &        $<$0.5  &    ...     &     ...    &   ...     &   ...      \\
TOL0109-383     &        $<$0.5  &    ...     &     ...    &   ...     &   ...      \\
I Zw 1          &        $<$0.4  &    ...     &     ...    &   ...     &   ...      \\
\noalign{\smallskip}
\hline
\end{tabular}
\end{flushleft}
1 = in km/s\\
2 = in arcseconds \\
3 = from Br\,$\beta$ (case B, see Hummer \& Storey 1987)\\
4 = from Br\,$\alpha$(see note 2)\\
5 = from Br\,$\gamma$ (Kowara et al. 1987, see note 2)\\
6 = from Br\,$\gamma$ (Goldader et al. 1995, see note 2)
}
\end{table}



 \begin{figure}
   \resizebox{\hsize}{!}{\includegraphics{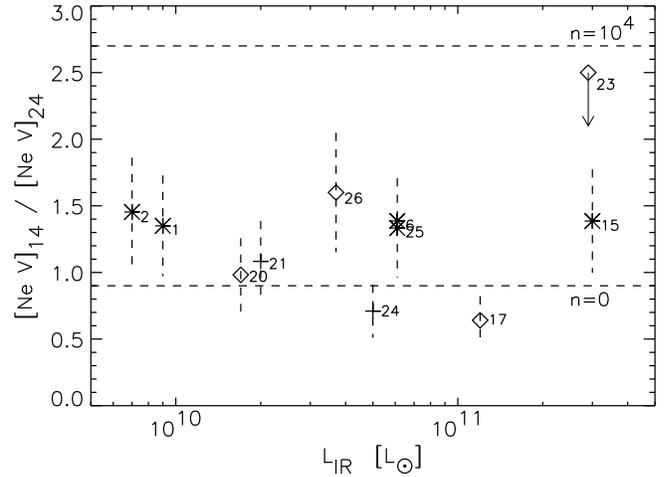}}
   \caption{The ratio of [Ne\,V] 14.3/24.3 $\mu$m. Diamonds: Seyfert 1s. Stars:
            Seyfert 2s, Plus signs: NLXGs. The theoretical low density limit ($\approx$0.9) 
            and the ratio for n$_e$=10$^4$cm$^{-3}$ (2.7), for a temperature of 10000K,
            are shown as dashed lines. Error bars for a 20\% line flux error are 
            indicated. The numbering of galaxies can be found in the first column of 
            Table \ref{tab:targets}.}
              \label{F:ne5ratios}
    \end{figure}

\section{Physical conditions in the NLR}
\label{s:nlr_cond}
Diagnostics of the physical conditions in the Narrow Line Region
are of interest given the wide range of possible excitation mechanisms
and interpretations for its emission. These include models assuming
ionization-bounded photoionization regions (e.g. T. Alexander et al. 1999, 2000),
more complex photoionization models invoking a mixture of ionization-bounded
and matter-bounded clouds (Binette et al. 1996, 1997) as well as shock
excitation, for example, through jet-cloud interaction (e.g. Axon et al. 1998). 
Traditional optical diagnostics have been
applied to large sets of objects but do have shortcomings. The
[S\,II] $\lambda\lambda$6716,6731 density diagnostic applies to low 
ionization species that may not be representative of the higher excitation 
regions and may be contaminated by non-AGN emission. The [O\,III]
$(\lambda 4959+\lambda 5007)/\lambda 4363$ temperature diagnostic 
includes a weak and blended line that is notoriously difficult
to measure. Combining infrared fine-structure lines with   
optical lines provides additional options. Given the large observing 
apertures for mid-IR instruments like ISO-SWS, their use should be 
restricted mainly to high 
excitation species that are not contaminated by spatially 
extended starburst emission. 

\subsection{Density}
\label{s:nlr_cond1}
Density-sensitive ratios of bright infrared fine-structure lines are observable
for a range of ionization potentials. Because of possible contamination by 
starburst emission (cf. Lutz et al. 2000b for the longer 
wavelength lines in NGC\,1068), we do 
not discuss [S\,III] 18.71/33.48$\mu$m and [Ne\,III] 15.55/36.01$\mu$m.  
The perhaps most interesting ratio is [Ne\,V] 14.3/24.3 $\mu$m.
These lines are not diluted by starburst contributions, since they have 
a lower ionization potential of 97eV and are undetected
in ISO spectra of starburst galaxies (Genzel et al. 1998). Their ratio
is largely insensitive to electron temperature and to 
extinction variations (see T. Alexander et al. 1999, their Fig. 3,
for the theoretical [Ne\,V] ratio as function of density).

All objects with measurable [Ne\,V] in our sample have [Ne\,V] ratios 
indicating a low density  
(n$_e$ = a few 100 to a few 1000 cm$^{-3}$), well below the critical densities
($\approx$ 5 x 10$^4$ and 5 x 10$^5$ cm$^{-3}$), see Fig. \ref{F:ne5ratios}. 
The average ratio is 1.1$\pm$0.4 for the (4) Seyfert 1s, and 1.3$\pm$0.3 for the (6) Seyfert 2s,
i.e. NLR densities for the two types agree for this fairly small
sample, consistent with unification. 


\begin{figure}
\resizebox{\hsize}{!}{\includegraphics{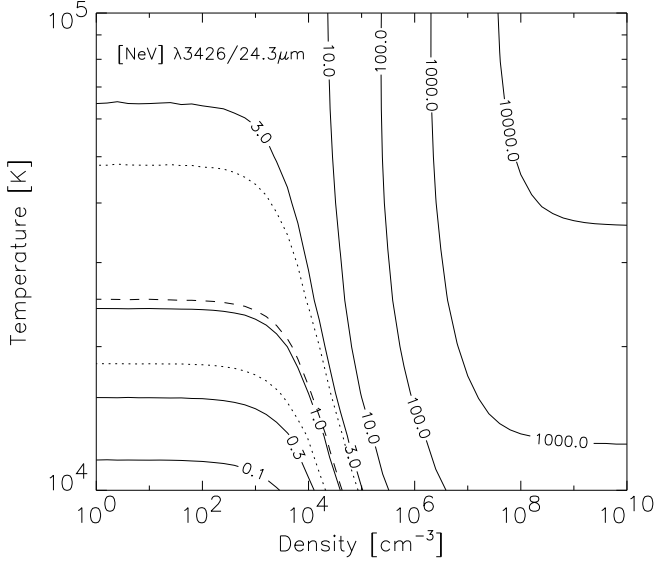}}
\caption{Temperature-sensitive ratio of the 3426\AA\ and 24.3$\mu$m lines
of [Ne\,V] as a function of electron density and temperature. The dashed
line indicates the ratio for NGC\,1068 and the dotted lines its uncertainty 
range considering measurement errors and uncertainty of extinction.}
\label{fig:ne5temp}

\end{figure}

\subsection{Temperature}
\label{s:nlr_cond2}
Temperature diagnostics for photoionized gas invoke a combination of 
transitions that originate in levels separated by at least several 1000K.
Combinations of ground state infrared fine-structure lines with optical/UV
forbidden lines of the same ion provide maximum leverage for this task 
and represent excellent diagnostics provided practical requirements
can be met: (1) Both IR and optical/UV lines have to be well observed,
with decent S/N and line-to-continuum. (2) Aperture corrections have to 
be understood - note the much larger ISO aperture compared to commonly used
optical apertures. (3) Contamination by non-NLR emission has to be low. (4)
Extinction corrections have to be understood. The last requirement
is not trivial to meet. The extinction
towards a NLR will not follow a simple foreground screen, as most
evidently shown by the presence of intra-NLR dust (e.g. Cameron et al. 1993),
and line profile variations between optical and infrared (cf. Lutz et al.
2000b for NGC 1068). Lines observed at different wavelengths may 
therefore not sample the same gas volume, even if fluxes are adjusted using
simple extinction corrections. Hence we ignore combinations UV/IR (e.g. [O\,IV],
[Mg\,V]) where these
problems are most severe, and discuss a number of optical/IR diagnostics
focussing on the well studied objects NGC 1068 and Circinus.  
Of particular interest are highly ionized regions not well sampled
by the optical [O\,III] or [N\,II] temperature diagnostics.

\begin{figure}
\resizebox{\hsize}{!}{\includegraphics{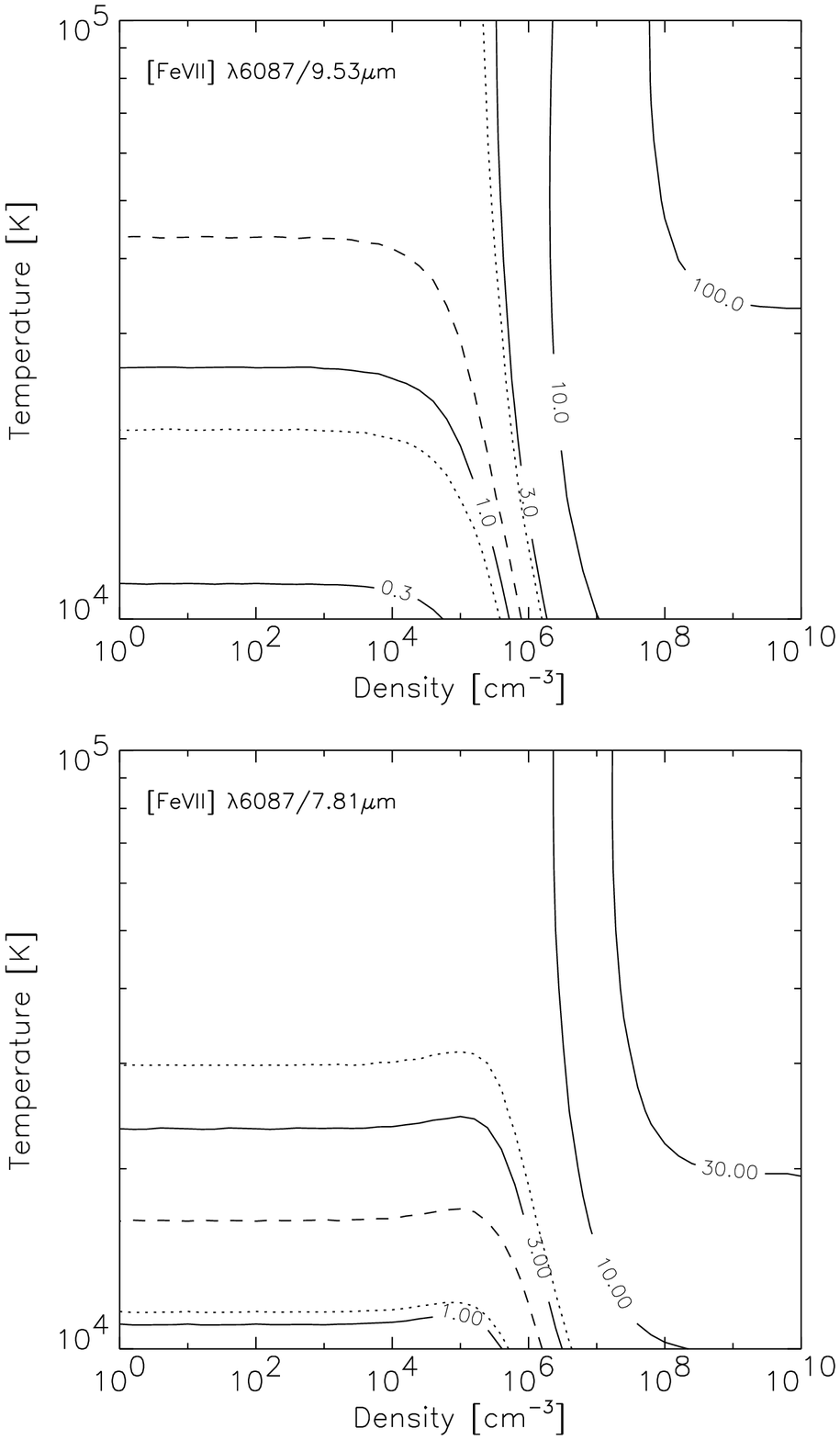}}
\caption{Temperature-sensitive ratio of the 6087\AA\ and the 9.53$\mu$m (top)
and 7.81$\mu$m (bottom) lines
of [Fe\,VII], as a function of electron density and temperature. The dashed
line indicates the ratio for NGC\,1068 and the dotted lines its uncertainty 
range considering measurement errors and uncertainty of extinction.}
\label{fig:fe7temp}
\end{figure}

One interesting species is [Ne\,V] (lower ionisation potential 97eV) which 
has very bright optical and mid-infrared transitions. Figure~\ref{fig:ne5temp} shows
the ratio predicted for the 3426\AA\ and 24.3$\mu$m lines, using the atomic data
of Lennon \& Burke (1994) and Nussbaumer \& Rusca (1979). The range of 
allowed values for NGC\,1068  (combining our data with optical data of Marconi
et al. (1996) and adopting A$_V$=0.71) is also indicated, adding a
20\% measurement error for each line, and a 0.25mag uncertainty in A$_V$. 
to produce the maximum deviation. 
These uncertainties (particularly the one on extinction) lead to a wide 
range of electron temperatures ($\sim$17000-45000K) being consistent with the 
NGC\,1068 data for its density of 2000\,cm$^{-1}$. 

Another high excitation species offers the potential of reducing the
extinction effects. The brightest optical line of [Fe\,VII]
is found at the longer wavelength of 6087\AA . With a lower ionisation 
potential of 99eV, [Fe\,VII] samples a region similar to
[Ne\,V]. The ratio of the optical [Fe\,VII] 6087\AA\ line 
to the mid-infrared [Fe\,VII] lines is little sensitive to electron 
density in the regime determined above, and
forms a diagnostic of electron temperature in the NLR. Figure~\ref{fig:fe7temp}
shows the ratios from solving the rate 
equation using the atomic data of Berrington et al. (2000, and priv. 
communication\footnote{Collision strengths for 10000K and corrections of three
erroneous transition probabilities}).
We have observed the mid-IR [Fe\,VII] lines in NGC\,1068 and Circinus. The
extinction corrected line ratios for NGC 1068 lead to very different electron 
temperatures: $\approx$43000K from 6087\AA/9.53$\mu$m but $\approx$16000K from
6087\AA/7.81$\mu$m. This discrepancy is also reflected in the extinction 
corrected ratio of the 9.53$\mu$m and 7.81$\mu$m lines which is 1.4, 
significantly different from the $\approx$3--3.5 expected for a wide range of 
conditions from the atomic data. For Circinus, the extinction corrected
ratio of the 9.53$\mu$m and 7.81$\mu$m lines is similarly low, about 1.1.
There is no immediate explanation for
this inconsistency between mid-IR FeVII fluxes that is observed independently
in two sources. An explanation by extinction uncertainties is unlikely,
since $\approx$10mag of additional visual extinction would be needed to
selectively weaken the 9.53$\mu$m line which is inside the silicate feature.
Uncertainties in the atomic data might be another possibility but are
only partially supported by observations of the same lines in the planetary 
nebula NGC 7027: Salas et al. (2001) observe a ratio of 1.84 for the two mid-IR
lines, closer to the value of 2.73 expected for the conditions in this nebula.
   
We summarize that, while there are indications for high (20000-30000K) electron 
temperatures in the part of the narrow line region sampled by E$_{\rm ion}\sim$100eV
species, there is considerable uncertainty on extinction corrections and
atomic data which prevents firm conclusions.

  \begin{figure}
   \resizebox{\hsize}{!}{\includegraphics{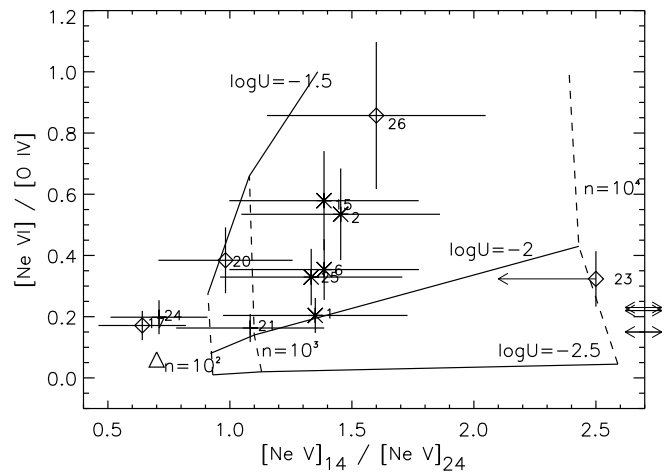}}
   \caption{Line ratio diagram of typical NLR tracers.
    Diamonds: Seyfert 1s. Stars: Seyfert 2s, Plus signs: NLXGs. 
    Numbering of galaxies as in Fig. \ref{F:ne5ratios}.
    The supernova remnant RCW103 (Oliva et al. 1999) is shown as a triangle.
    The [Ne\,VI]/[O\,IV] values for three additional sources are indicated as horizontal
    arrows on the right hand y-axis (from top to bottom: 3C120, Mkn509, Mkn573).
    To constrain ionization parameter and density we have overplotted a model
    grid by Spinoglio et al. (2000) for a power law ionizing continuum of index
    $\alpha$ = -1.0: solid lines are log U = -2.5, -2, and -1.5,
    dashed lines are for n = 10$^2$, 10$^3$, and 10$^4$.
    }
              \label{F:sedmodeling}%
    \end{figure}

\begin{table*}
\caption[]{\label{tab:profiles} Classes of line profile asymmetries}
\begin{flushleft}
\begin{tabular}{lllll}
\hline\noalign{\smallskip}
\rule[-2mm]{0mm}{2mm}   & Class I                &  Class II  &               Class IIIa   & Class IIIb   \\ \hline     
optical [O\,III] profile & symmetric              & asymmetric              & asymmetric               & asymmetric  \\
MIR [O\,IV] profile      & symmetric, =optical    & symmetric               & asymmetric, =optical     & asym./non-gauss., $\neq$optical  \\
examples                & 3C120$^1$, PKS2048$^2$ & Mkn509$^2$, NGC5643$^2$ & NGC3783$^2$,NGC4151,$^3$ & NGC1068$^3$,Mkn1$^1$, Mkn3$^3$,  \\
                        &                        &                         & NGC7469$^1$, Tol0109$^2$ & Mkn463$^2$, NGC5506$^3$   \\
\noalign{\smallskip}
\hline
\end{tabular}
\end{flushleft}
1 = optical line profile from Vrtilek \& Carleton 1985\\
2 = optical line profile from Whittle 1985\\
3 = optical line profile from Veilleux 1991\\
\end{table*}

\subsection{Excitation}
\label{s:central_source}

Varying relative line strengths of low and high ionizaton lines, e.g. of the Neon sequence,
seem to indicate a large variation in excitation among the galaxies in our sample.
Line ratios from ions in different stages of excitation along with a photoionization model can 
be used to reconstruct the NLR radiation field produced by
the central ionizing source. 
Mid-Infrared lines are particularly suited because they are little sensitive to extinction
and electron temperature, and because they span a wide range of ionization potentials.
In detailed studies we have analyzed three nearby Seyfert nuclei of our sample with elaborate
photoionization models:
Circinus (Seyfert 2 plus starburst), NGC4151 (Seyfert 1.5) and NGC1068 (Seyfert 2), see Moorwood et al. (1996), and T. Alexander et al. (1999, 2000).
These models were able to reconstruct the intrinsic spectral energy distribution (SED) of the 
ionizing source
in the extreme UV, where a `Big UV Bump' 
around 100eV is expected as a 
signature from a hot thin accretion disk. The results are consistent with such a `Big UV Bump', but also suggest for some sources of both types (Seyfert 1 and 2) the presence of neutral absorbers between the AGN's extreme ultraviolet emitting source and the NLR. Such (UV) absorbers have been suggested independently by studies of UV absorption lines (Kriss et al. 1992, Kraemer et al. 1999, 2001). 
This detailed photoionization modeling requires a large number of mid-IR lines (of good S/N) supplemented with UV/optical/near-IR lines. For most of the galaxies in our sample the compilation of lines from ISO and the literature is not as complete as for the three examples given above. Hence, we do not attempt 
a similar modeling for them. 
Instead, we try to constrain the NLR excitation by directly comparing our observations to 
standard photoionization models from the literature.

The main input parameters for photoionization models are the electron density n$_e$ 
and the ionization parameter U, i.e. the number of ionizing photons per hydrogen 
atom at the inner face of the ionized cloud. The ratio of lines from ions of similar 
ionization potential but with different critical densities (e.g. the [Ne\,V] 14/24 ratio) 
are good tracers of the electron densities (see Sect. \ref{s:nlr_cond1}). 
Vice versa, the ratio of lines with similar 
critical density but from ions of different ionization potential (like [Ne\,VI]/[O\,IV]) 
is sensitive to the ionization parameter. In Fig. \ref{F:sedmodeling} we have used this to construct 
a diagram to constrain n$_e$ and U. We have chosen the lines of [Ne\,V], [Ne\,VI] and [O\,IV] 
because they are not affected by photoionization by stars and because they are generally 
among the brightest high ionization lines. 
For comparison we show in Fig. \ref{F:sedmodeling} 
the location of the supernova remnant RCW103
(Oliva et al. 1999), as an example for a strong shock source, and 
a photoionization model grid taken from Spinoglio 
et al. (2000). This model grid was computed for a power law with an ionizing continuum 
of index $\alpha$=-1.0 and for various ionization parameters and electron densities.
We draw two conclusions from this comparison: firstly, 
all galaxies are consistent with these standard photoionization models, with average ionization 
parameters log U between -1.5 and -2 and (as seen already in Sect. \ref{s:nlr_cond1}) 
average densities between a few 100 and a few 1000 cm$^{-3}$). 
Such a simple comparison can not, however, distinguish between
simple power laws and more complex `Big UV Bump' models. As noted earlier, the detailed modeling
of single sources required for such an analysis is outside the scope of this paper. 
Secondly, within our (small) sample, we do not see significant differences
between the AGN sub-types. 

\subsection{NLR line profiles}
\label{s:profiles}

Structures and 
velocity fields in the NLR can be studied by an analysis of  
line profiles. 
Numerous studies of the emission line profiles in Seyfert galaxies have shown that
in many cases the lines exhibit blueward asymmetries and are blueshifted with respect to
the galaxies' systemic velocity. 
The common interpretation for these profile asymmetries  is that they are caused by differential 
extinction in an outflow or inflow of clouds with a modest amount of mixed-in dust.
Observations in the infrared can obviously test these scenarios, because infrared lines
suffer more than an order of magnitude less extinction than in the optical. Hence, they
should not or only marginally show asymmetries.
Sturm et al. (1999) and Lutz et al. (2000b) have presented the first such studies of
optical-to-mid-IR line profile comparisons for NGC 4151 and NGC 1068.
In these two sources the outflow-plus-dust scenario 
seems to be (at least partially) wrong, since the
MIR lines show asymmetries similar to their optical counterparts.

\begin{figure}
\resizebox{\hsize}{!}{\includegraphics{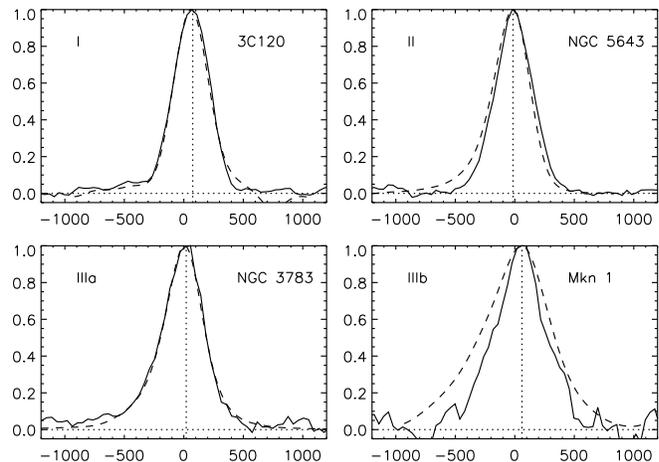}}
\caption{Comparisons between optical and MIR 
line profiles (continuum subtracted and normalized to peak flux density), 
for four different cases as discussed in the text. Solid lines show
the MIR [O\,IV] 26$\mu$m line, taken from the ISO-SWS data set. Dashed lines are optical
[O\,III]5007{\AA} lines, smoothed to the SWS resolution. References for the optical 
line profiles are given in Table \ref{tab:profiles}. The velocity scale on 
the x-axis (in [km/s]) is relative to the systemic velocities given in 
Table \ref{tab:targets}.}
\label{fig:profiles}
\end{figure}

Optical [O\,III] line profile information exists for many objects in our data set. 
These can be used for a comparison with the MIR [O\,IV] line, as
described in Sturm et al. (1999). We have convolved the optical line profiles with
the SWS instrumental profile (Gauss profiles with FWHM as given in 
Table \ref{tab:fluxes-fine}) in order to smooth these profiles to the resolution of
SWS. Four such examples of an optical-MIR comparison are shown in Fig. \ref{fig:profiles}.
The objects in our sample can be grouped roughly into three different classes in terms
of line profile asymmetries and agreement of MIR lines with their optical counterparts, 
as summarized in Table \ref{tab:profiles}. 
In some cases (Class I) the optical line profile of [O\,III] is quite symmetric and of 
gaussian shape. In these cases, not surprisingly, the MIR and optical line profiles 
match each other very well. Another group of objects (Class II) shows strong
blueward asymmetries in the optical, while the MIR lines are rather symmetric. 
This is exactly what
scenarios with infall our outflow of NLR clouds with mixed-in dust predict.
We only found two such cases in our sample. The third class of objects has asymmetric
optical lines, but MIR lines which are inconsistent with these scenarios. 
This class can be further
divided into two sub-classes: in Class IIIa the MIR lines are asymmetric, too, and
agree well with the optical lines. This case has been studied in more detail in an
analysis of NGC 4151 by Sturm et al. (1999), and can be explained, for instance, by
a true asymmetry in the distribution of the NLR clouds, or, in the case of NGC 4151,
by a central, optically very thick, but geometrically thin absorber on parsec scales. 
In Class IIIb the NIR lines are asymmetric (or symmetric but with non-gaussian profiles), 
but different from their optical counterparts. One member of this class is NGC 1068,
which has been analyzed by Lutz et al. (2000b). 
This suggests that parts of the NLR are significantly obscured in
      the optical, but not enough to also block the MIR lines. Similar to Class
      IIIa, the remaining MIR profile asymmetries may be either due to an 
      intrinsic asymmetry of the NLR, or due to a very high density obscuring
      component which is hiding part of the NLR even from infrared view.
We note that for some of the objects with MIR lines of good S/N 
(Cen A, M 51, MKN 573, NGC 1365, and NGC 7582)
there is,
to our knowledge, no 
(suitable) optical line profile information available in the literature.  

Many objects in our sample exhibit differences in the profiles of lines with
different ionization potential. For instance,
the Circinus galaxy shows symmetric low 
ionization lines, but asymmetric
high ionization lines with the typical blueward asymmetries. Vice versa, in NGC 7582 
high ionization lines are very symmetric, while the low ionization lines are strongly 
asymmetric. Correlations, as well as anti-correlations, 
of line asymmetries with critical density and/or ionization potential
have been claimed in many publications. For NGC 7582 a 
contribution to the low ionization lines from a starburst component with asymmetric 
spatial distribution could be an additional/alternative solution.  
For the Circinus galaxy the situation is even more complex, since Oliva et al. (1994) 
reported asymmetric low ionization lines.
Finally, for some objects in our sample all MIR line profiles appear to be quite
similar. For instance, the profiles of Cen A, M 51, MKN 573, NGC 1365 (for which no optical
counterparts exist) are quite symmetric and of gaussian shape, regardless of ionization
potential or critical density. 

It appears that there is no unique answer to what causes the line profile 
asymmetries in Seyfert galaxies. Extinction by dust on different spatial scales
and with varying column densities, 
sub-structure and true asymmetries in the spatial distribution of NLR clouds may all 
play a role with varying degrees of importance.
For some objects the assignment to a certain group is not unique and depends also on
S/N. We use this classification purely to obtain an overview of this complex issue. 
We refrain from a more detailed analysis of line profiles in this paper because it requires
a careful study of the effects of aperture differences (spatial resolutions) between
the optical and MIR observations, and, in many cases, a larger data set of high
ionization lines with good S/N.


  \begin{figure}
   \resizebox{\hsize}{!}{\includegraphics{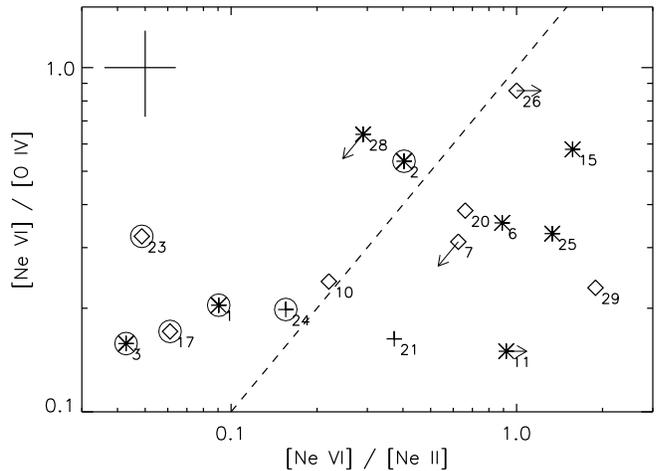}}
   \caption{The ratio [Ne\,VI]/[O\,IV] vs. [Ne\,VI]/[Ne\,II]. Same symbols and numbering
            as in Fig. \ref{F:ne5ratios}.
            A typical error bar (according to a 20\% individual line flux error) is shown in the upper left                                     corner. An empirical dividing line is drawn as dashed line. Composite sources according to
their PAH spectra are encircled. For galaxies \#26, 28 and 29 no ISO PAH spectra exist.}
              \label{F:ne6o4_ne6ne2}%
    \end{figure}


\section{Composite sources}
\label{s:star_contrib}

As discussed in Sects. \ref{s:introduction} and \ref{s:targets}
for many active galaxies an additional contribution from star forming regions within the 
SWS apertures is to be expected. In these cases (`composite sources') 
some fraction of the 
measured fluxes of low lying fine structure
lines (excitation potential $\le$ 50 eV) will be produced by photoionization from stars 
rather than from the AGN. 
In the previous sections we have therefore concentrated on high excitation lines 
([O\,IV], [Ne\,V], [Ne\,VI], [Fe\,VII])
with little or no contamination from possible starburst components. However, for our considerations in the following sections low ionization lines like [Ne\,II] will be important. Hence, an independent identification
of composite sources is desirable.

  \begin{figure}
   \resizebox{\hsize}{!}{\includegraphics{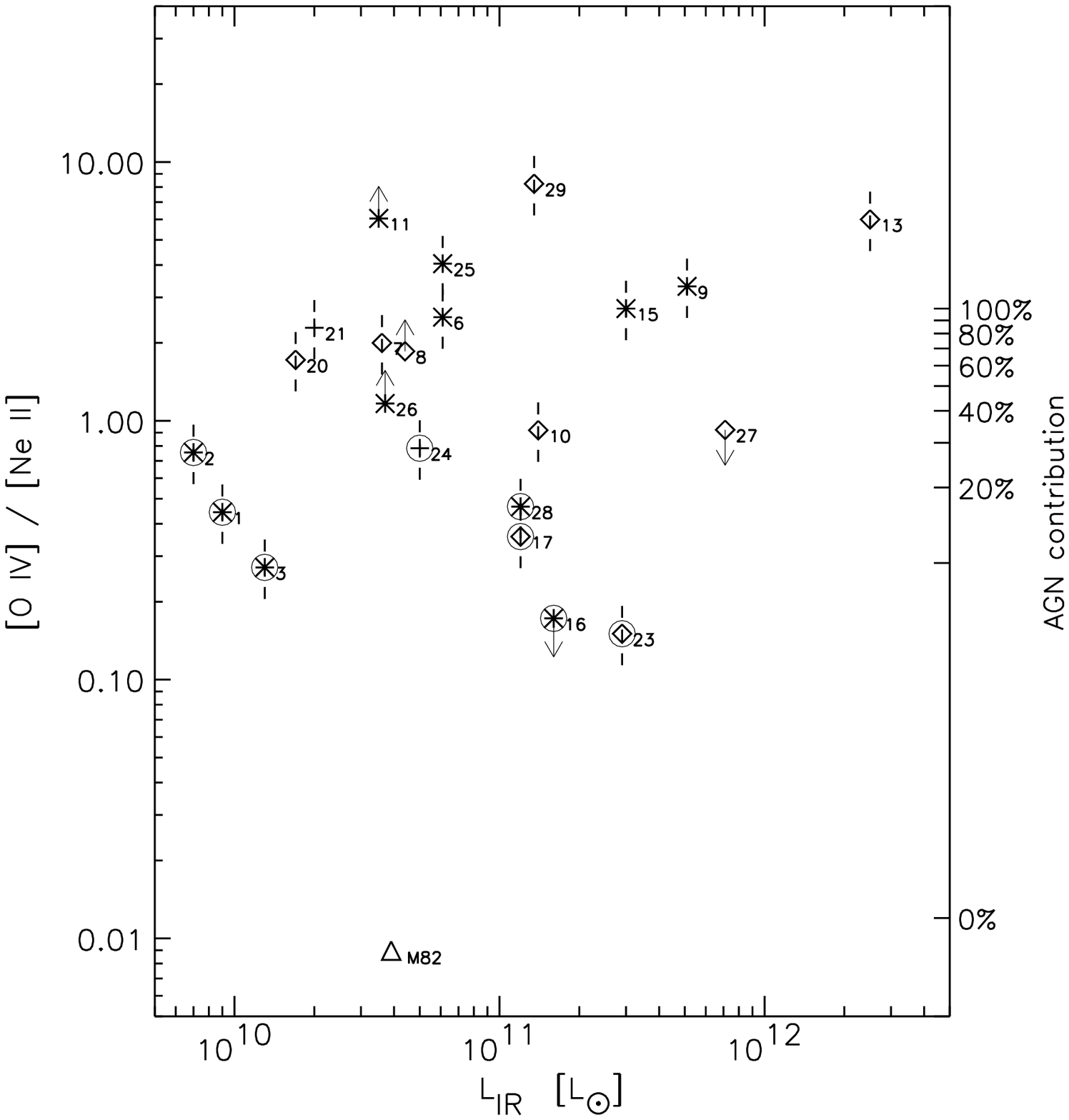}}
   \caption{The ratio [O\,IV]25.9$\mu$m / [Ne\,II]12.8$\mu$m. Same symbols and numbering
            as in Fig. \ref{F:ne5ratios}.
            Composite objects are encircled.
            The starburst prototype M 82 is shown as a triangle. Denoted on the right hand
            y-axis is a simple linear `mixing' model of AGN contribution to the bolometric 
            luminosity, in percent (see text). The [O\,IV]/[Ne\,II] ratios for 100\% AGN and 
			0\% AGN (=100\% starburst) are assumed to be 2.7 and 0.012 respectively. }
              \label{F:o4vsne2}%

    \end{figure}

One such independent indicator is the
strength of PAH features in the 5 to 12 $\mu$m range, which
are known to be strong in star forming regions, but more or less absent in the spectra 
of pure AGNs. Genzel et al. (1998), Rigopoulou et al. (1999), and Tran et al. (2001)
have analysed a large number of low resolution ISOPHOT-S,
ISOCAM-CVF and SWS spectra and used the line-to-continuum (L/C) ratio of the
7.7$\mu$m PAH feature to distinguish between starburst and AGN dominated sources. 
Following their method (and using the ISO archive as well as the spectra published in
Rigopoulou et al. 1999 and Spoon et al. 2002) we classify Cen A, Circinus, M 51, NGC 613, NGC 1365, 
NGC 7469, and NGC 7582, i.e. those sources with a 7.7 $\mu$m L/C ratio 
greater than 1, as composite sources.
As a consistency check, and
for the cases where no PAH spectrum is available (\mbox{Tol 0109}, \mbox{I Zw 92}, and \mbox{3C 120)}, 
we have plotted in Fig. \ref{F:ne6o4_ne6ne2}
the position of the targets in a diagram of [Ne\,VI]/[O\,IV] versus [Ne\,VI]/[Ne\,II].
As discussed in Sect. \ref{s:central_source}, the ratio [Ne\,VI]/[O\,IV] traces 
the AGN excitation without contamination from a possible star formation component.  
[Ne\,VI]/[Ne\,II] on the other hand, which is also an excitation indicator, can be influenced 
by contributions from star forming regions to the [Ne\,II] line. In such a diagram one expects
the pure AGNs to lie roughly along a diagonal (both ratios increase with increasing 
hardness of the radiation). Composite sources, however, will have relatively stronger [Ne\,II] lines 
and should lie in a region further to the left than the pure AGN sources. 
In fact, all sources with known substantial starburst contributions, i.e. with 
strong PAH features, fall onto the left part of the diagram (Cen A, Circinus, M51, NGC1365, NGC 7469, and
NGC7582, marked with an enclosing circle in Fig. \ref{F:ne6o4_ne6ne2}). 
Pure AGNs (like NGC4151 and NGC1068) are located in the right hand part. 
In Fig. \ref{F:ne6o4_ne6ne2} we have indicated an empirical dividing line (dashed). For simplicity 
we have chosen the line where [O\,IV] equals [Ne\,II].  
According
to this diagram one additional source without (good S/N) PAH data, I Zw 92 (\#28), 
has to be classified as composite 
source. This classification is confirmed by independent UV to near-IR observations (Heckman et al. 1997, 
Gonzalez Delgardo et al. 2001). 
We consider Mkn 509 (\#10) a borderline case.
Another source, NGC1275, can not be shown at its exact position in the diagram, but according to the upper limits for [O\,IV] and [Ne\,VI] (and the measured value for [Ne\,II])
the position must be in the composite source region.
The galaxies that we classify as composite sources according to this scheme
are indicated in Table \ref{tab:targets} as `Sy + SB'.
The location of
composite sources (as indicated by strong PAH features) in the left
part of Fig. \ref{F:ne6o4_ne6ne2} suggests that a small ratio of [NeVI] and [NeII] 
\mbox{(\raisebox{-0.5ex}{$< \atop \widetilde{}$} 0.1)} 
is an independent empirical indicator for a composite object.

  \begin{figure}
   \resizebox{\hsize}{!}{\includegraphics{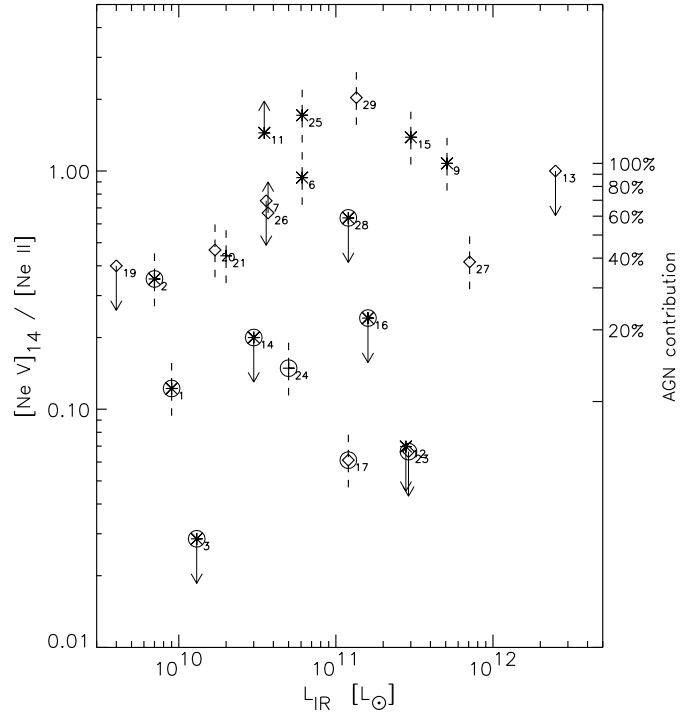}}
   \caption{The ratio [Ne\,V]14.3$\mu$m / [Ne\,II]12.8$\mu$m. Same symbols and numbering
            as in Fig. \ref{F:ne5ratios}.
            Composite objects are encircled.  
            Denoted on the right hand
            y-axis is a simple linear `mixing' model of AGN contribution to the bolometric luminosity, 
            in percent (see text). 
			The [Ne\,V]/[Ne\,II] ratios for 100\% AGN and 0\% AGN (=100\% starburst) 
			are assumed to be =1.1 and 0 respectively. }
              \label{F:ne5vsne2}%
    \end{figure}
%


\section{New mid-IR diagnostic diagrams}
\label{s:new_diag}

Deep surveys carried out by future infrared missions (such as SIRTF, SOFIA, ASTRO-F,
or Herschel) will sample infrared bright galaxies over a
wide range of redshifts and luminosities.
Quantitative spectroscopy of mid-infrared emission lines will
be an important diagnostic tool for determining the detailed properties of 
distant, dusty galaxies, the source of the extragalactic background, and 
the origin of nuclear activity in galaxies.
Many of
the fundamental questions of galaxy formation and evolution depend
substantially on the fraction of the total energy output of distant
sources that is produced by star formation rather than AGN activity.
A large
energy contribution from hidden AGNs would complicate the deduction of
the star formation history of the Universe from galaxy luminosity
functions. 
In the past, optical line ratios have been used as a tool 
for distinguishing between the 
different possible excitation 
mechanisms and energy sources in galaxies (e.g Baldwin et al. 1981; 
Veilleux \& Osterbrock 1987). This works very
well for objects with dust extinction less than A$_{\rm V}$$\sim$~5, but
becomes unreliable for heavily obscured objects (see, e.g., Veilleux et
al. 1995; Veilleux et al. 1999). Hence, mid-infrared analogues of the classical optical
diagrams are highly desirable.
  \begin{figure}
   \resizebox{\hsize}{!}{\includegraphics{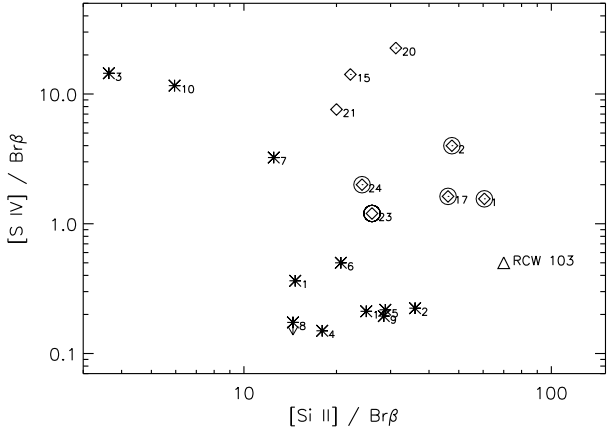}}
   \caption{A MIR diagnostic diagram to distinguish starbursts and AGNs in
        dusty galaxies -- here: [S\,IV] 10$\mu$m/Br$\beta$ versus [Si\,II] 34$\mu$m/Br$\beta$. 
        AGNs are shown as diamonds, starburst galaxies (Verma et al., in prep.)
        as stars. The shock source RCW 103 is shown as triangle. Composite objects are 
        encircled. For three AGNs Br$\beta$ was 
        calculated from other Brackett lines (see Table \ref{tab:fluxes-recomb}). 
        The numbering of the AGNs is taken from Table \ref{tab:targets}, for the starbursts it is as follows:
        1:M82, 2:IC342, 3:IIZw40, 4:NGC253, 5:NGC3256, 6:NGC3690A, 7:NGC4038/39, 8:NGC4945,
        9:NGC5236, 10:NGC5253, 11:NGC7552.}
              \label{F:s4_si2}%

    \end{figure}
%
  \begin{figure}
   \resizebox{\hsize}{!}{\includegraphics{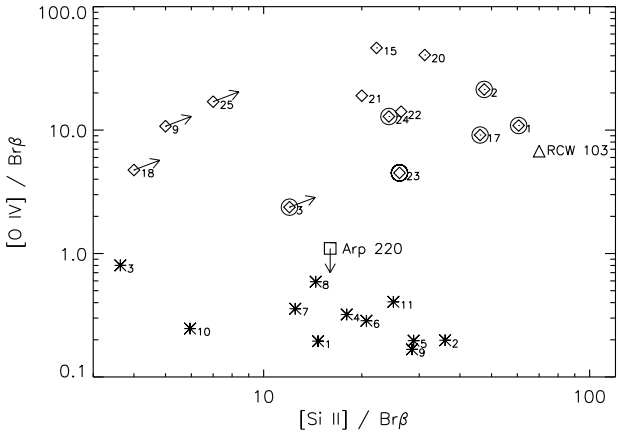}}
                 \caption{Another example of MIR diagnostic diagrams -- here: 
            [O\,IV] 26$\mu$m/Br$\beta$ versus [Si\,II] 34$\mu$m/Br$\beta$.
            Symbols and numbering as in Fig. \ref{F:s4_si2}. The ULIRG Arp 220 
            (Sturm et al. 1996) is shown as a square, the SN remnant RCW 103 
            (Oliva et al. 1999) as triangle.}
            \label{F:o4_si2}%
    \end{figure}
%
  \begin{figure}
   \resizebox{\hsize}{!}{\includegraphics{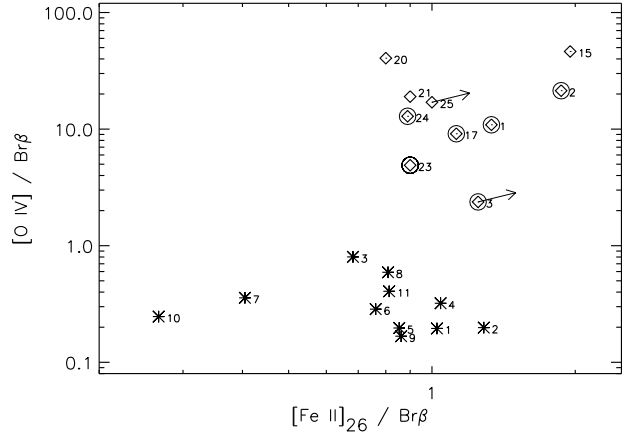}}
   \caption{Another example of MIR diagnostic diagrams -- here: 
            [O\,IV] 26$\mu$m/Br$\beta$ versus [Fe\,II] 26$\mu$m/Br$\beta$.
            Symbols and numbering as in Fig. \ref{F:s4_si2}.}
              \label{F:o4_fe2}%
    \end{figure}
Genzel et al. (1998) presented a first empirical version of such a mid-IR tool, a diagram
with the flux ratio of a high to a low excitation line ([O\,IV]/[Ne\,II], or [Ne\,V]/[Ne\,II])
on one axis together with the AIB (PAH) strength on the other axis. Their finding that the [O\,IV]/[Ne\,II] 
ratio is much higher in AGNs than in starbursts can now be confirmed on a broader statistical basis.
In starburst galaxies this ratio reaches values of a few times 0.01 at most (Lutz et al. 1998), 
whereas in the AGNs of our sample it does
not drop below 0.1 (1.0 for pure AGNs), as shown in Fig. \ref{F:o4vsne2}. For comparison we show in Fig.
\ref{F:ne5vsne2} the ratio of [Ne\,V]14$\mu$m/[Ne\,II].
Those active galaxies in our sample with the lowest ratios in Figs. \ref{F:o4vsne2} and \ref{F:ne5vsne2}
are the ones we have identified earlier as composite sources (encircled in the figures), 
i.e. AGNs with a significant contribution from 
star formation to their spectra.
On the right hand y-axes of Figs. \ref{F:o4vsne2} and \ref{F:ne5vsne2}
we have denoted a simple linear `mixing' model to read off the
AGN contribution to the bolometric luminosity. This model will be further discussed in 
Sect. \ref{s:fir_source}.


Perhaps the best optical 
discriminator between photoionization by power-law spectra and by OB stars
is the diagram in Fig. 3 of Veilleux \& Osterbrock (1987). It employs a 
moderately ionized line ([O\,III]) on one axis and a line of [O\,I], normalized
to hydrogen recombination lines.
Our data set allows the construction of MIR analogues of this `VO diagram' which are
much less prone to extinction. 
%
We have selected
[S\,IV]10.5$\mu$m instead of [O\,III], which has an almost identical ionization potential,
[Si\,II]34$\mu$m instead of [O\,I], i.e. a line with an
ionization potential below 13.6 eV, and Br\,$\beta$ for normalization. 
For comparison we have used the ISO-SWS starburst data set of Verma et al. 
(in prep.). Starbursts and AGNs are clearly separated in this diagram (Fig. \ref{F:s4_si2}), 
not only
because [S\,IV] is stronger in AGNs than in normal starbursts, but also because [Si\,II] 
is 
(on average) stronger, due to the higher importance of partially ionized 
zones in AGNs. Low metallicity starbursts, like IIZw40 and NGC5253, have much harder radiation fields
than normal starburst galaxies. In these cases [S\,IV] can be as strong as in AGNs, but [Si\,II] 
is much weaker. The overall appearance of Fig. \ref{F:s4_si2} is thus very similar to the optical
VO diagram. In contrast to its optical counterpart our mid-IR version also includes Seyfert 1
galaxies, since we did not detect any broad line components in the Br\,$\beta$ lines.
For a complete comparison of the optical and mid-IR 
diagrams our data set is missing galaxies of the LINER type. Their position in the mid-IR diagram
is expected in the lower right corner, but remains to be tested in future infrared missions.  

Several other versions of mid-IR diagnostic diagrams (with lines at different wavelengths)
are similarly well 
suited for a distinction of excitation mechanisms. This allows 
to cover different redshift ranges or to adjust the method to the wavelength coverage
of different detectors.
For instance, it is possible to replace 
[S\,IV]10.5$\mu$m with [O\,IV]26$\mu$m, 
and [Si\,II]34$\mu$m with [Fe\,II]26$\mu$m (Figs. \ref{F:o4_si2} and \ref{F:o4_fe2}).
Similar diagnostic diagrams, based on theoretical modelling, with different sets of 
(mostly weaker) lines
have been proposed by Spinoglio \& Malkan (1992) and Voit (1992).
In contrast to the optical VO diagrams, the mid-IR versions can be applied to
dusty systems with much higher extinctions, such as Ultraluminous Infrared Galaxies (ULIRGs). 
ULIRGs are believed to be local analogues of those distant dusty galaxies, which are (and will be) 
found in great numbers in deep infrared galaxy surveys. As discussed above, the identification
of their energy source will be a major task for future infrared missions.
In Fig. \ref{F:o4_si2} we show the position of the well known ULIRG Arp 220 
(Sturm et al. 1996). 
Its position in the diagram is consistent with predominant powering
by intense star formation, which is in accordance with earlier studies 
(e.g. Sturm et al. 1996, Genzel et al. 1998).




\section{The source of the far-infrared luminosity} 
\label{s:fir_source}

In the previous section we have presented tools to detect AGNs in dusty galaxies.
The presence of an AGN, however, does not automatically imply
that the AGN activity is the dominant energy source in these galaxies.
A significant fraction of the bolometric luminosity of Seyfert galaxies
is emitted in the mid and far infrared by thermal (dust) emission.
For a quantitative understanding of the contribution from AGNs
to the total energy output of distant galaxies 
it is therefore essential to know 
whether the
source powering the IR emission is (re-)radiated light from 
the AGN or from distributed star formation in the host galaxy.


In the following we will compare luminosities in different tracers for 
Seyferts and Starbursts  
in an attempt to shed light on these issues.
We have selected [O\,IV] (as a high excitation line), [Ne\,II] (as a low excitation
line), and the mid- and far-infrared continua. 
In all cases, we will 
keep track of the composite AGNs already identified in Sect. \ref{s:star_contrib}.

The  mid-IR ($\lambda < 40\mu$m) emission of AGNs is dominated by 
dust continuum re-radiating energy from black hole accretion. Orientation 
effects in a unified scheme cause variations in its level between Seyfert types
1 and 2, and extended star formation can contribute, more visibly in Sy2s 
with their lower AGN continuum (Clavel et al. 2000). In contrast, the origin of
the 40-120$\mu$m 
far-infrared emission of AGNs is considerably uncertain. Both 
the AGN proper and star formation have been advocated as energy source in the
literature (cf. the review by Genzel \& Cesarsky 2000 and references
therein). For starbursts, the mid-IR emission will be dominated by
the aromatic emission features and transiently heated small grains. The
far-IR emission originates in large dust grains, and all are heated by newly 
formed or older stars (e.g. Desert, Boulanger \& Puget. 1990). 

[O\,IV] originates purely from
the NLR in AGNs. In a unified scheme, the NLR line luminosity should
be orientation independent and a good tracer of AGN power, in particular
when using an extinction insensitive and modest excitation line like [O\,IV].  
The {\em much} fainter starburst [O\,IV] emission
relates to shocks or some very hot stars (Lutz et al. 1998). [Ne\,II]
is a pure and fairly good tracer of the hot star emission in starbursts. In AGNs
the [Ne\,II] from the NLR is more easily contaminated by  starburst emission
than the higher excitation [O\,IV] line.
   
\begin{figure}
   \resizebox{\hsize}{!}{\includegraphics{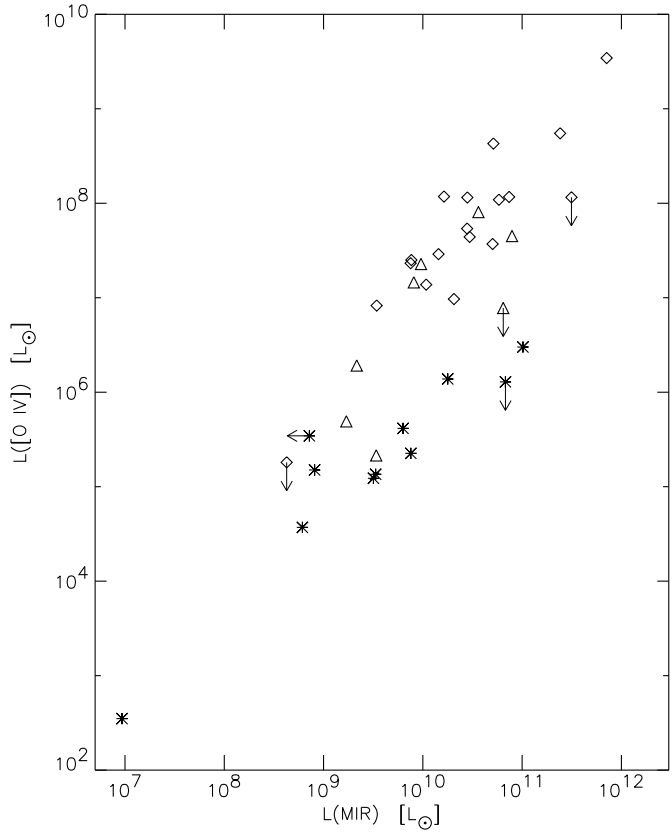}}
   \caption{[O\,IV] luminosity vs. MIR luminosity for AGNs (diamonds) and starburst 
            galaxies (stars). Composite sources (see text) are indicated as triangles.}
              \label{F:o4_mir}%
    \end{figure}

\begin{figure}
   \resizebox{\hsize}{!}{\includegraphics{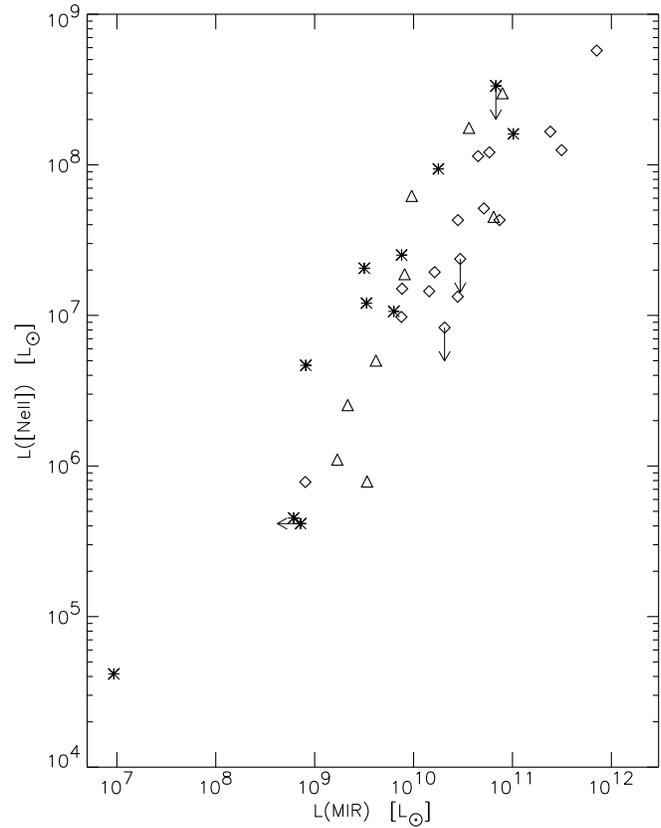}}
   \caption{[Ne\,II] luminosity vs. MIR luminosity for AGNs (diamonds) and starburst 
            galaxies (stars). Composite sources (see text) are indicated as triangles.}
              \label{F:ne2_mir}%
    \end{figure}

\begin{figure}
   \resizebox{\hsize}{!}{\includegraphics{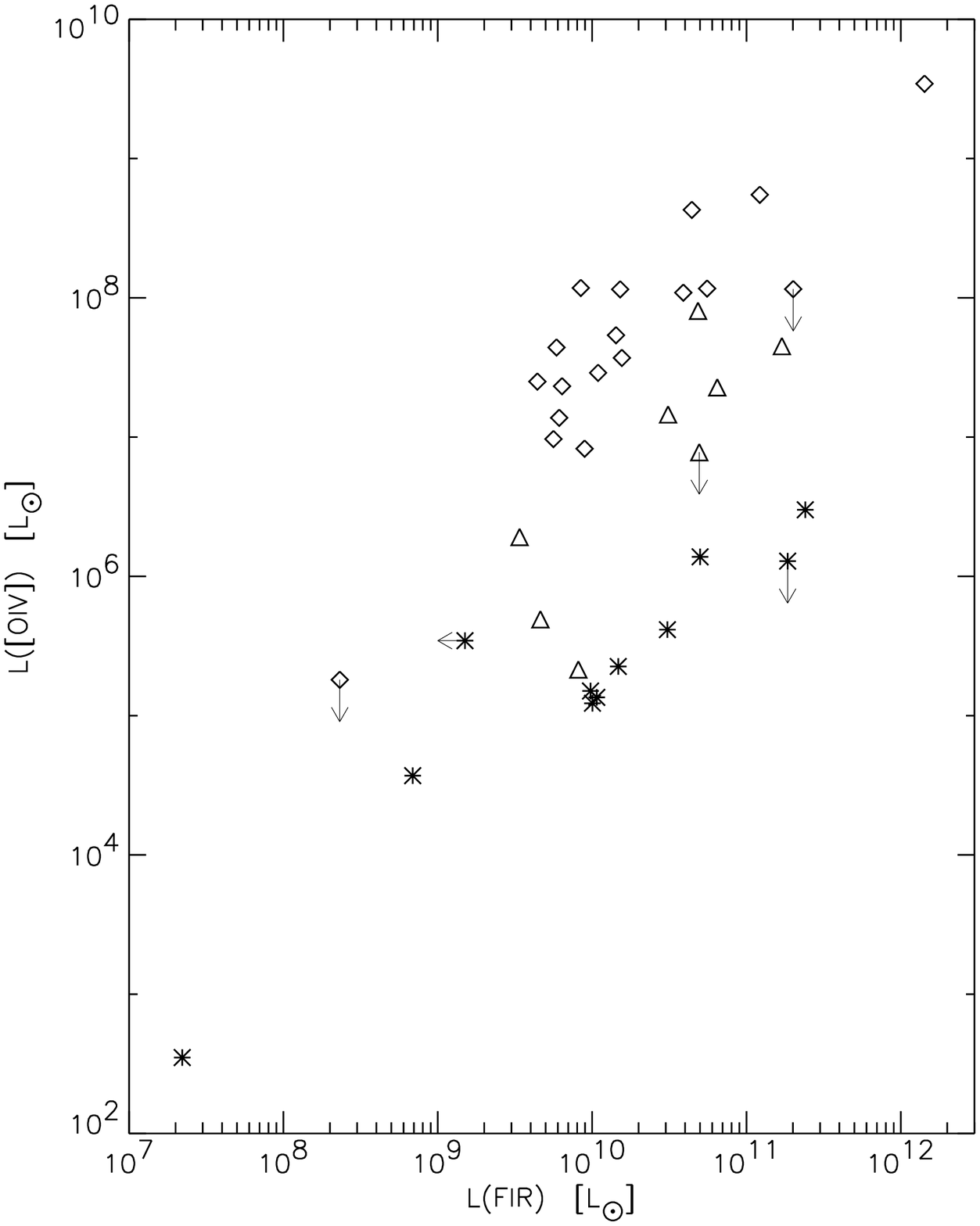}}
   \caption{[O\,IV] luminosity vs. FIR luminosity for AGNs (diamonds) and starburst 
            galaxies (stars). Composite sources (see text) are indicated as triangles.}
              \label{F:o4_fir}%
    \end{figure}

\begin{figure}
   \resizebox{\hsize}{!}{\includegraphics{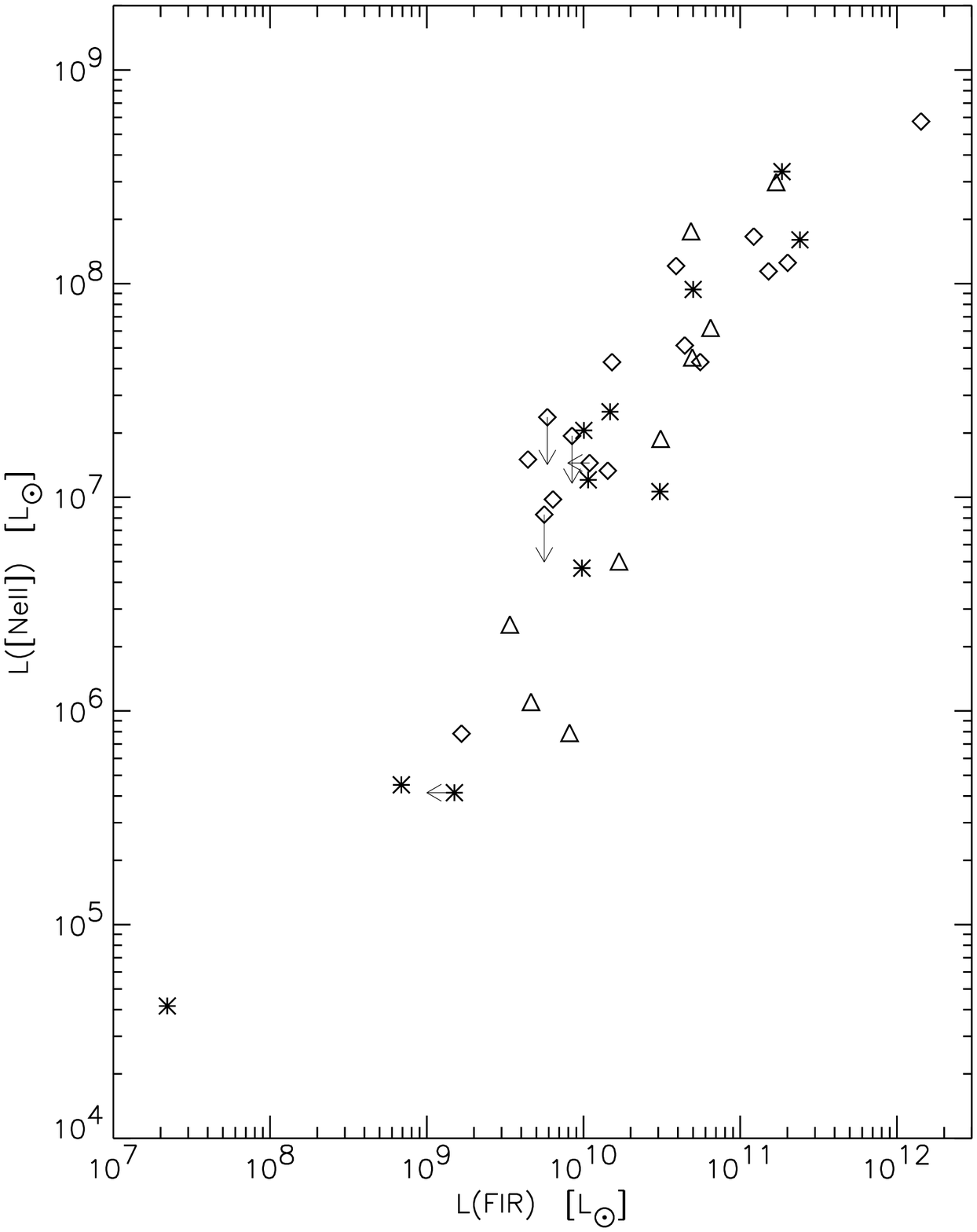}}
   \caption{[Ne\,II] luminosity vs. FIR luminosity for AGNs (diamonds) and starburst 
            galaxies (stars). Composite sources (see text) are indicated as triangles.}
              \label{F:ne2_fir}%
   \end{figure}

We have computed mid-IR (12+25$\mu$m) and far-IR (60+100$\mu$m) band luminosities from the respective 
IRAS fluxes following 
Sanders \& Mirabel (1996), using
\mbox{F$_{\rm MIR}$ = 1.26$\times$(5.16S$_{25}$+13.48S$_{12}$)$\times$10$^{-18}$ W/cm$^2$,} 
and 
\mbox{F$_{\rm FIR}$ = 1.26$\times$(S$_{100}$+2.58S$_{60}$)$\times$10$^{-18}$ W/cm$^2$,} 
where S is the flux density in Jy (from IRAS FSC), and 
\mbox{L=4$\pi$D$^2$F = 3.132$\times$10$^4\times$D$_{\rm Mpc}^2$F L$_{\odot}$.}
Our results are summarized in Figs. \ref{F:o4_mir}-\ref{F:ne2_fir}, 
showing correlations of [O\,IV] and
[Ne\,II] line luminosities with MIR and FIR continuum luminosities, for
pure Seyferts, composite Seyferts (both from this paper) and starbursts 
(from Verma et al. in prep.). The object classes are discriminated by 
plot symbols. For each object 
class separately, good correlations are observed (see also Prieto et al. 
2002 showing a similar analysis for a much smaller Seyfert sample and using 
ISOPHOT fluxes). The scatter is partly intrinsic but will also include a
component due to aperture differences between SWS lines and IRAS continua. 
This is irrelevant for the AGN NLRs which are smaller than the SWS aperture
but may affect some of the more extended starburst emission.  
The slopes are near 1 suggesting that the spectral 
properties of each object class do not change much over the overall luminosity 
range probed by our sample. Comparing different object classes, both 
Fig. \ref{F:o4_mir} and Fig. \ref{F:o4_fir} clearly reflect the much stronger [O\,IV] emission of 
Seyferts compared to starbursts (Genzel et al. 1998, Lutz et al. 1998).
The composite Seyferts tend to fall between pure Seyferts and starbursts,
again confirming our classification based on PAH emission, and consistent
with the fine structure line ratios (Sects. \ref{s:star_contrib}/\ref{s:new_diag}).
As seen in Fig. \ref{F:ne2_fir} and already noted by Genzel et al. (1998), the ratio of 
[Ne\,II] and FIR luminosity is very similar in Starbursts and AGNs. The 
difference in [NeII]/MIR between Seyferts and starbursts is somewhat
larger (Fig. \ref{F:ne2_mir}), due to the moderately MIR-brighter AGN SEDs 
(e.g., de Grijp et al. 1985, 1987). The small variations in the ratio of 
[Ne\,II] and infrared luminosities make this line a useful reference line for 
comparison of line ratios between different types of galaxies (see below).
     
\begin{figure}
   \resizebox{\hsize}{!}{\includegraphics{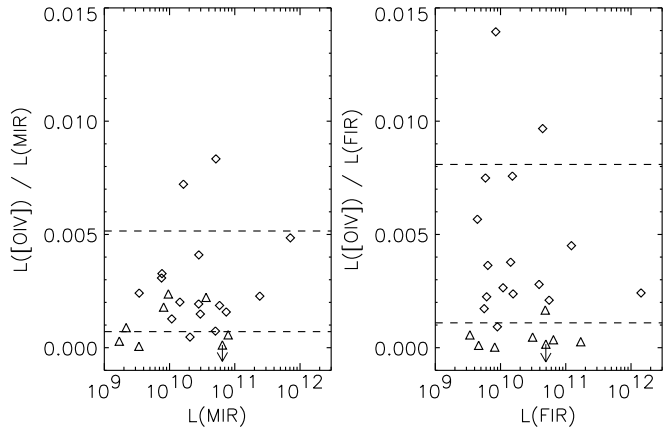}}
   \caption{[O\,IV] luminosity vs. MIR luminosity (left) and vs. FIR luminosity (right)
            for pure AGNs (diamonds) and composite sources (triangles).
            The $\pm1\sigma$ range around the average luminosity ratios (computed without composite sources)
            is marked with dashed lines.}
   \label{F:scatter}%
  \end{figure}

\begin{figure*}
   \centering
   \includegraphics{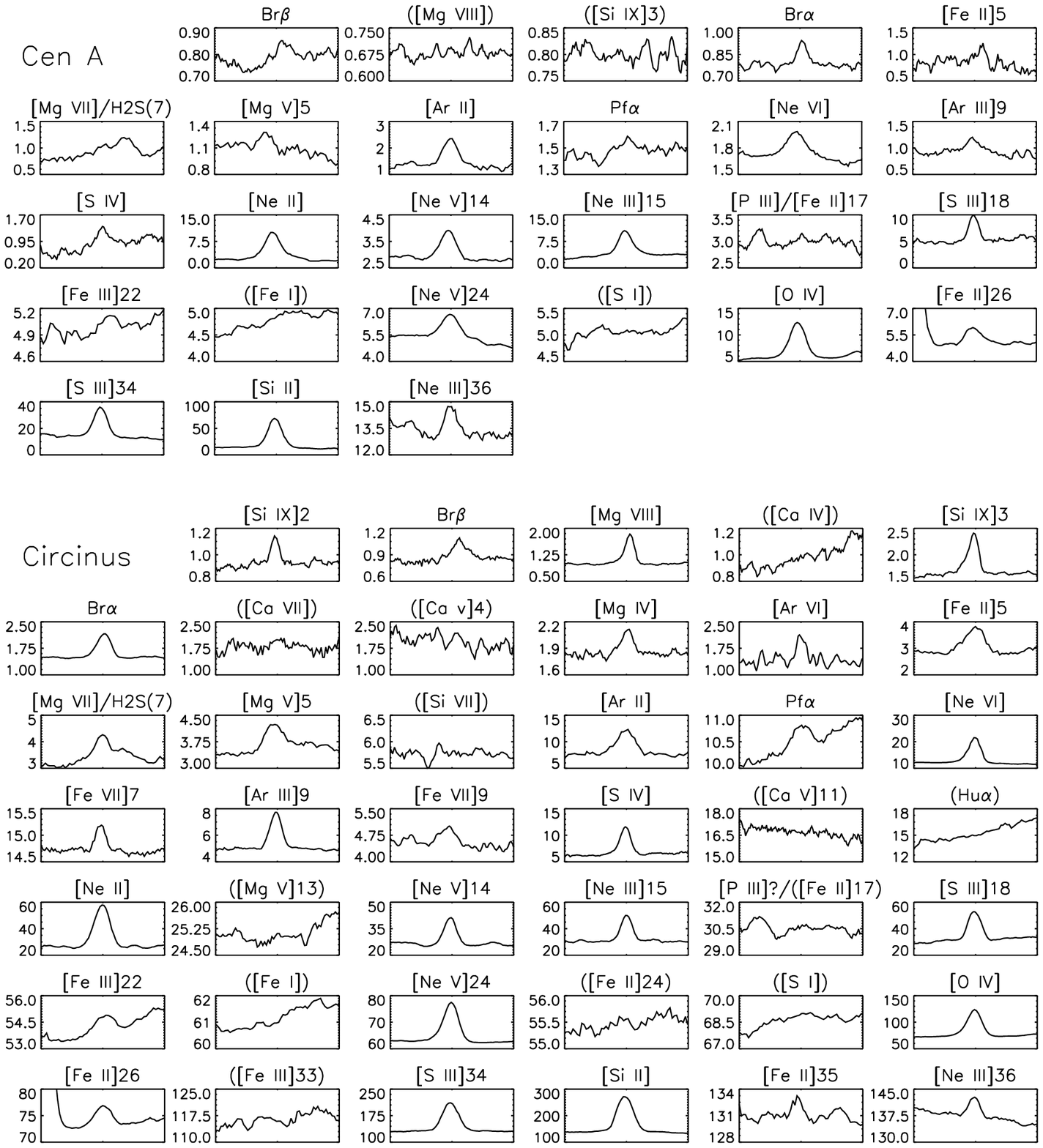}
   \caption{ISO-SWS spectra of Centaurus A and Circinus. Flux densities in [Jy] for a velocity range of 
             $\pm$1200km/s around the respective systemic velocity (v$_{\rm sys}$=cz, see Table \ref{tab:targets}).
             Identifications in brackets denote undetected lines with upper flux limits only.}
              \label{F:figspec1}%
    \end{figure*}

\begin{figure*}
   \centering
   \includegraphics{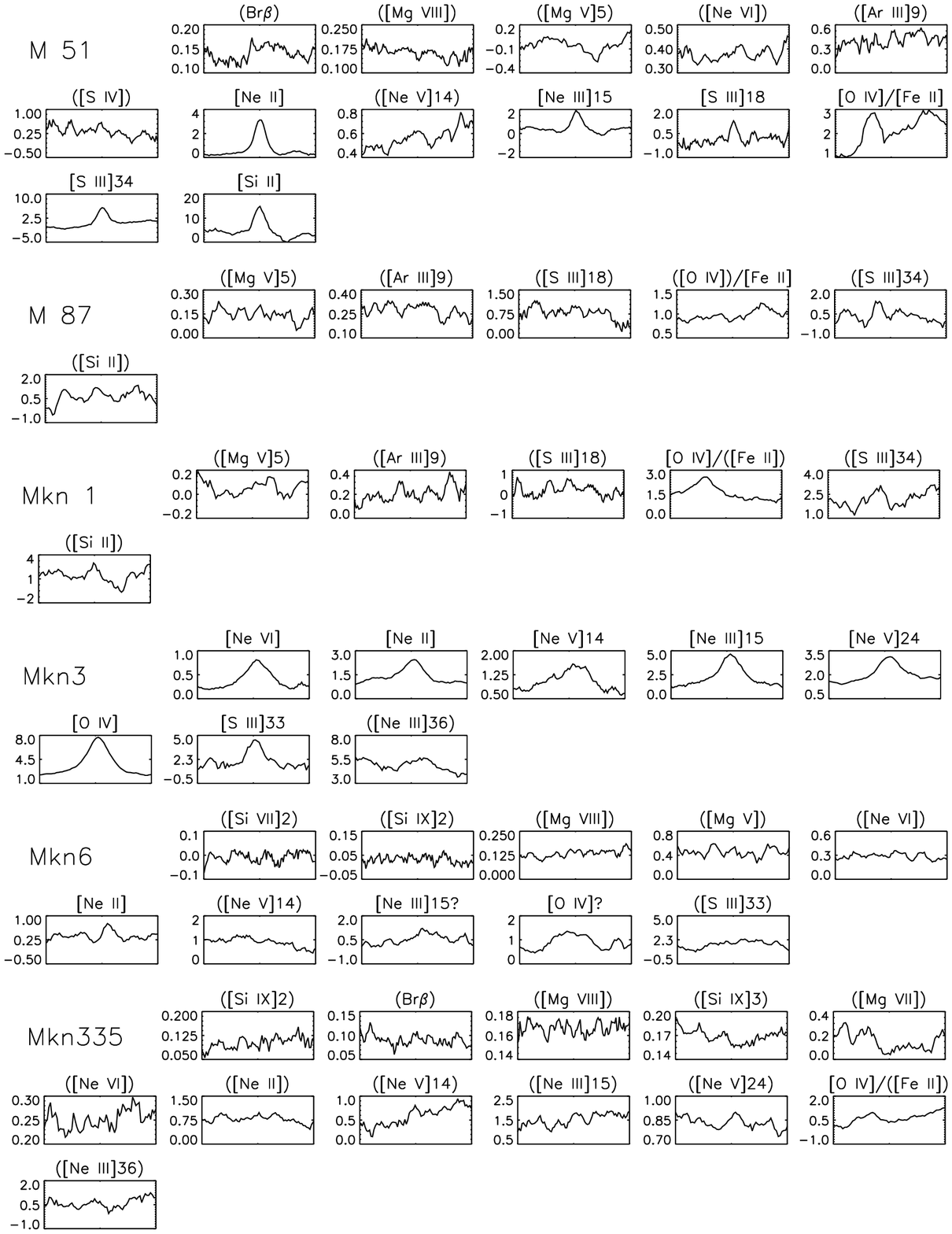}
   \caption{Further SWS spectra. Same units as in Fig. \ref{F:figspec1}. The range for the combined
             [O\,IV]/[Fe\,II] spectra is [-700,+1700]km/s around [O\,IV].}
              \label{F:figspec2}
    \end{figure*}

\begin{figure*}
   \centering
   \includegraphics{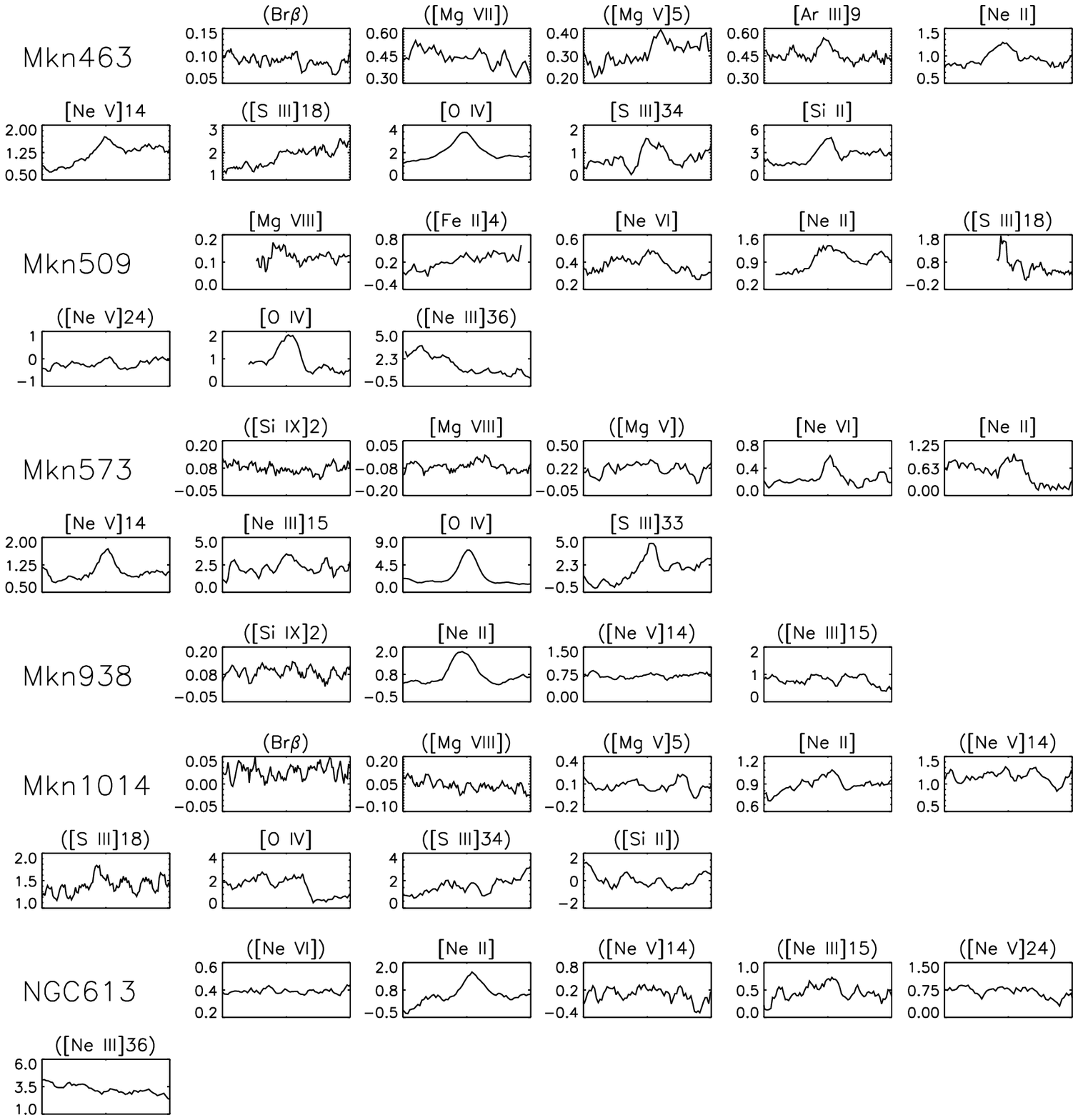}
   \caption{Further SWS spectra. Same units as in Fig. \ref{F:figspec1}.}
              \label{F:figspec3}
  \end{figure*}

\begin{figure*}
   \centering
   \includegraphics{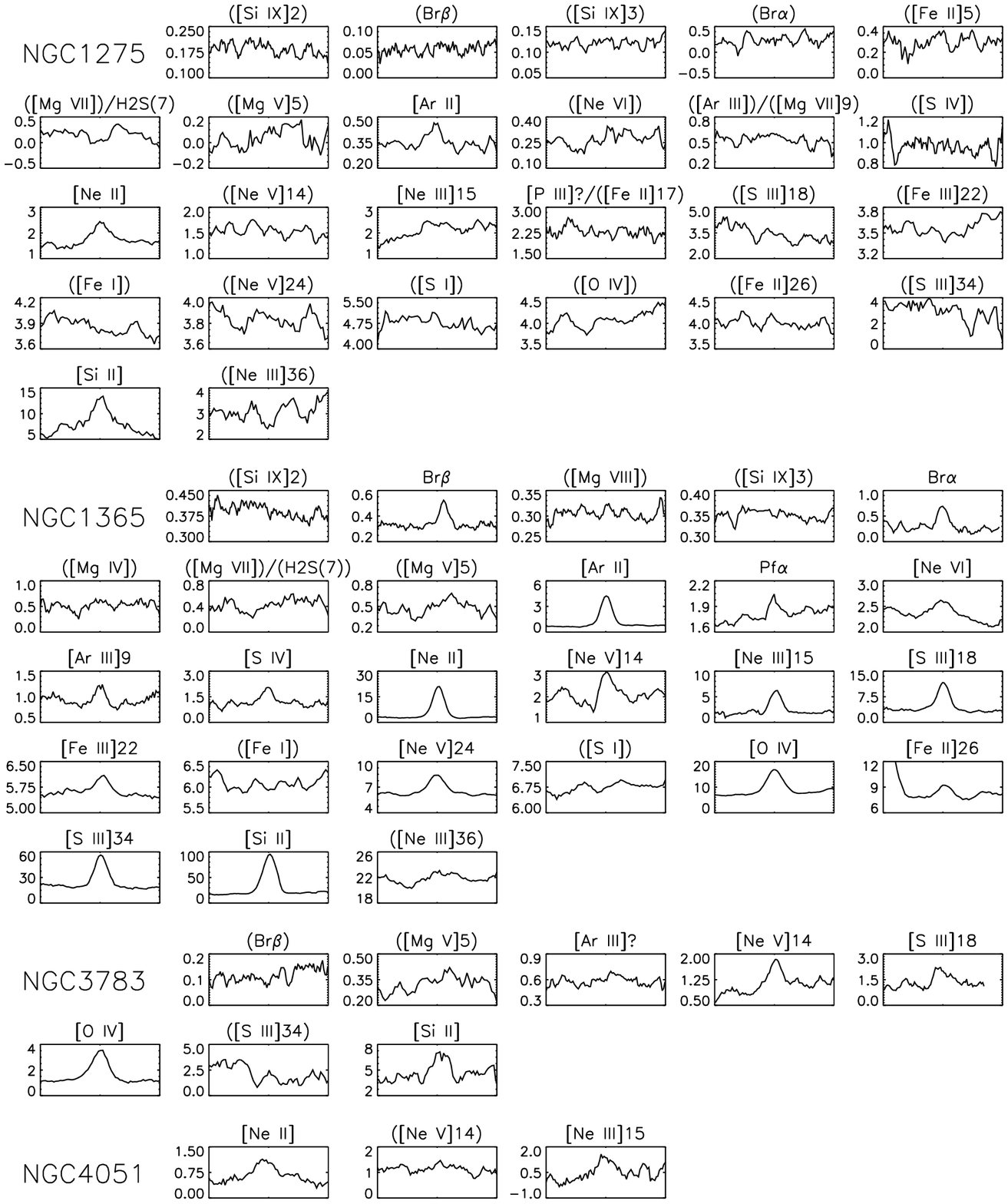}
   \caption{Further SWS spectra. Same units as in Fig. \ref{F:figspec1}.}
              \label{F:figspec4}
  \end{figure*}

\begin{figure*}
   \centering
   \includegraphics{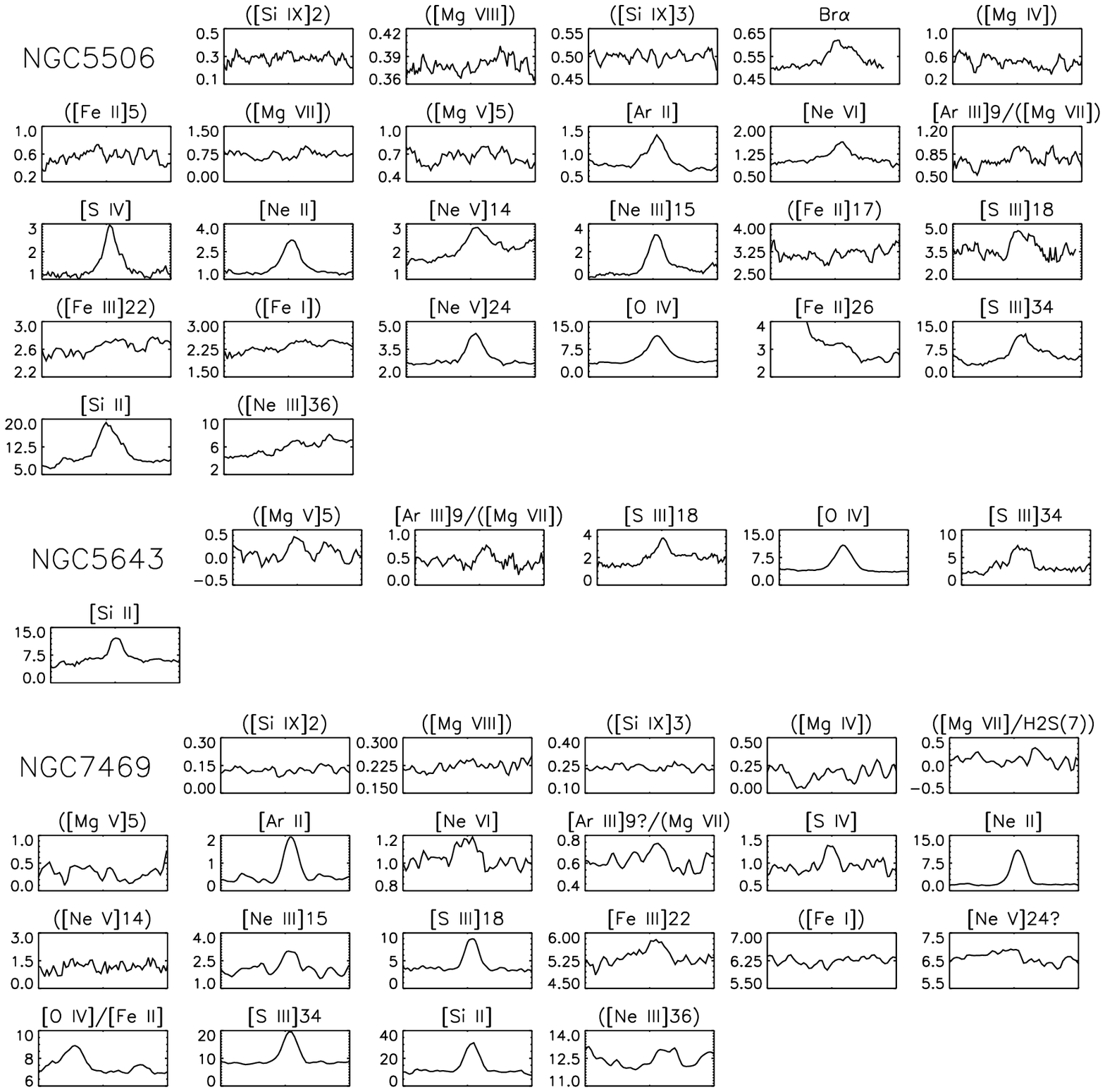}
   \caption{Further SWS spectra. Same units as in Fig. \ref{F:figspec1}. The range for the combined
             [O\,IV]/[Fe\,II] spectrum is [-700,+1700]km/s around [O\,IV].}
              \label{F:figspec5}
  \end{figure*}

\begin{figure*}
   \centering
   \includegraphics{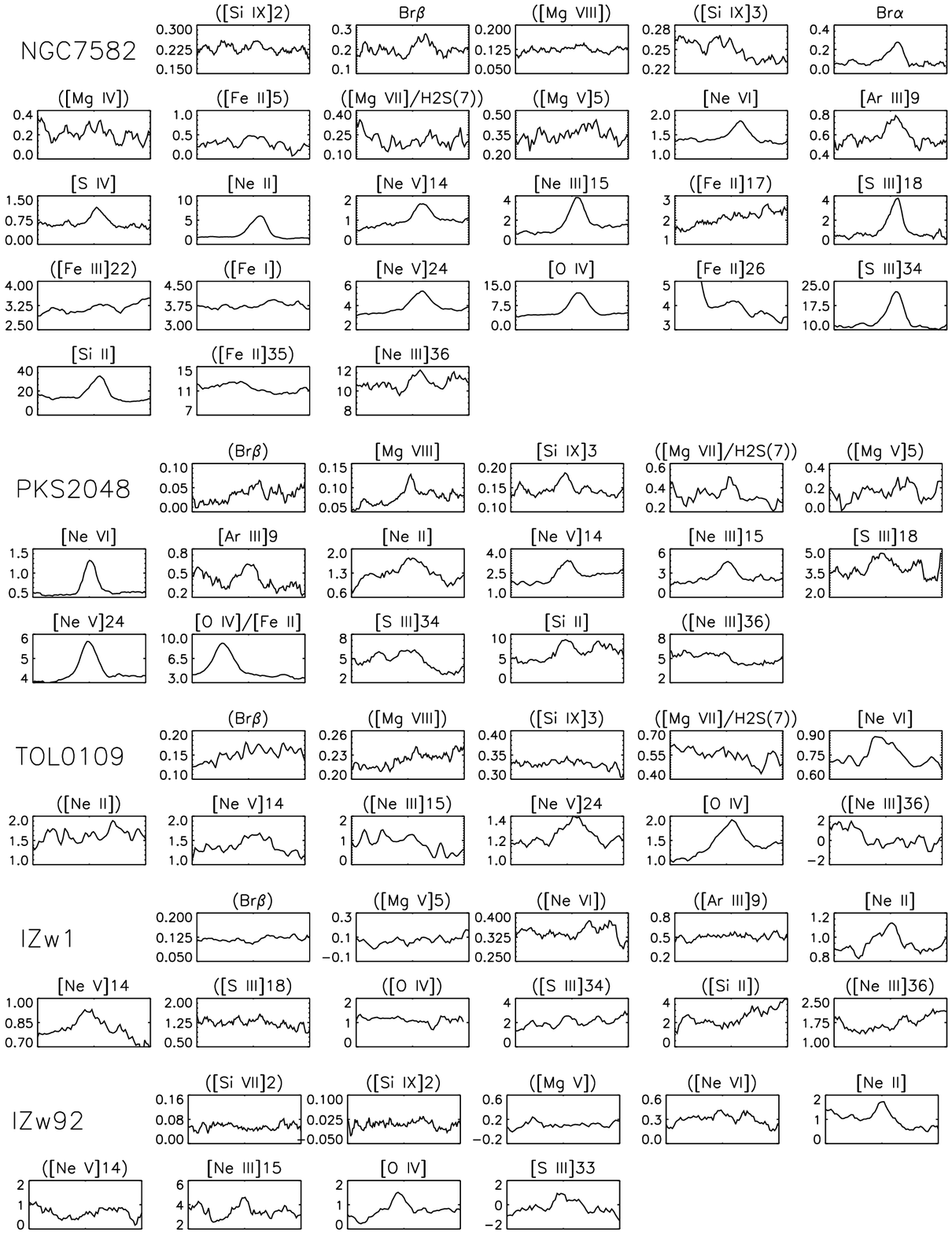}
   \caption{Further SWS spectra. Same units as in Fig. \ref{F:figspec1}.}
              \label{F:figspec6}
  \end{figure*}

\begin{figure*}
   \centering
   \includegraphics{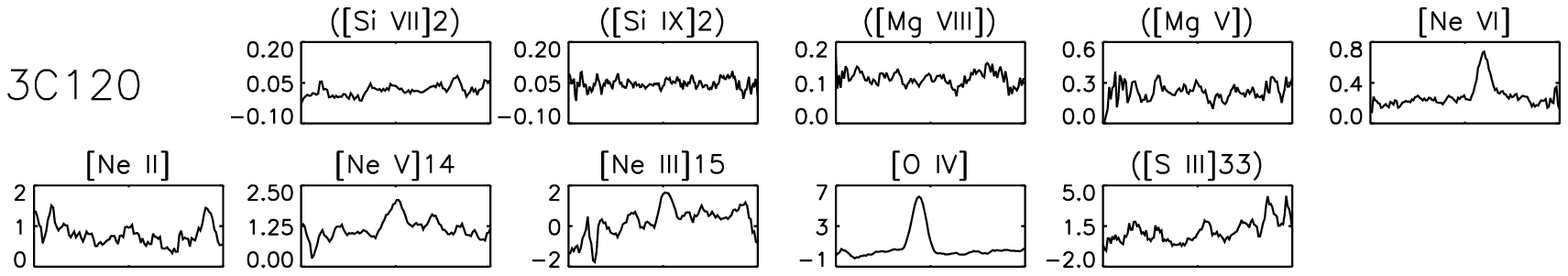}
   \caption{Further SWS spectra. Same units as in Fig. \ref{F:figspec1}.}
              \label{F:figspec7}
  \end{figure*}

A possible test for the origin of the AGN far-infrared emission is to
compare the quality of the correlations between [O\,IV] (assumed to trace
AGN luminosity) and MIR/FIR. If the AGN MIR emission were AGN heated
but not the FIR emission, one would expect less scatter in the
ratio [O\,IV]/MIR than in [O\,IV]/FIR. Figure \ref{F:scatter} 
indeed suggests the correlation with mid-infrared emission to be better,
but the effect is not strong (see also the similar conclusion of Prieto 
et al. 2002). Partly, the small magnitude of the effect may be simply due 
to the scatter in the 'reference' ratio [O\,IV]/MIR.
Even if [O\,IV] were a perfect measure of AGN power (which it clearly is not),
the orientation effects and larger scale contributions to the mid-IR
AGN emission (Clavel et al. 2000) would cause some scatter in the
left panel of Fig. \ref{F:scatter}. 
These results are consistent with a scenario where the mid-infrared luminosity
in AGNs is dominated by the active nucleus, whereas the far-infrared luminosity
is mainly produced by star formation. The considerable correlation between 
[O\,IV] and far-infrared luminosities in Seyferts (Fig. \ref{F:o4_fir}) would then simply
reflect 
a starburst-AGN connection in which starburst power and AGN power are linked. There is 
plenty of evidence reported in the literature for such a connection. For instance there
could be a feedback mechanism that controls (circum-)nuclear star formation and Black Hole
accretion rate. A more trivial link could simply be the common gas consumption habits
of starbursts and AGNs (see e.g. Cid Fernandes et al., 2001).
On the other hand, we can not exclude other scenarios in which, for example, 
the FIR luminosity of AGNs
is powered by the active nuclei.
The relation of the coronal line fluxes to the mid/far infrared
continuum fluxes does not allow us directly to quantify the AGN contribution to the 
total luminosity of their host galaxies with sufficient accuracy.
However, the constancy of the [Ne\,II]-to-FIR ratio can be used to
construct models for the AGN/starburst contribution to L$_{Bol}$ from the
ratios of MIR lines to [Ne\,II].
In Figs. \ref{F:o4vsne2} and \ref{F:ne5vsne2} (right hand y-axes) 
we show a simple linear mixing model, using the respective median ratios of
[O\,IV]/[Ne\,II] and [Ne\,V]/[Ne\,II] for
the pure AGNs and the pure starbursts in our sample (see figure caption). 
Those sources that we have identified as composite sources by means of their
PAH spectrum (or their position in Fig. \ref{F:ne6o4_ne6ne2}) lie in the 
10 to 50 percent domain of these mixing curves.   
When applying such simple models 
to individual sources one should keep in mind that there is some
scatter in the underlying relations, and the derived contributions in percent should not
be taken too literally. For statistical studies, however, these models provide a 
valuable means of quantifying the AGN contribution to the luminosity of dusty,
composite objects.

\section{Summary}
\label{s:summary}

We have presented an inventory of mid-infrared (MIR) lines detected in medium resolution (R$\sim$1500) 
ISO-SWS 2.4--45$\mu$m spectra of a sample of 29 galaxies with active nuclei. This data set
is rich in fine structure emission lines tracing the narrow line regions and 
(circum-)nuclear star 
formation regions, and it 
provides a coherent spectroscopic reference for future extragalactic studies in the mid-infrared.
From the [Ne\,V] 14/24 $\mu$m line ratios we conclude that all targets in our sample
with measurable [Ne\,V] emission 
have relatively low NLR electron densities (a few 100 to a few 1000 cm$^{-3}$).
Electron temperatures are difficult to derive.
There are indications for high (20000-30000K) electron 
temperatures in the part of the narrow line region sampled by E$_{\rm ion}\sim$100eV
species, but there is considerable uncertainty on extinction corrections and
atomic data which prevents firm conclusions.

A comparison of the observations to photoionization models showed that
all objects are consistent with standard photoionization models, i.e. the
spectral energy distribution of their central ionizing source can be explained
by a normal power law. These simple comparisons can not, however, distinguish between
simple power laws and more complex models of a `Big UV Bump' and possibly absorption. 
The detailed modeling
required for such an analysis is outside the scope of this paper. 
Within the statistical 
limitations of our sample we do not see any significant differences of NLR densities
or ionization between the different AGN types. 

The MIR lines in our data set show various degrees of profile asymmetries and
different levels of similarities to their optical counterparts. 
It appears that the cause for the line profile 
asymmetries in Seyfert galaxies is a complex mixture of extinction by dust on different 
spatial scales, sub-structure and true asymmetries in the spatial distribution of NLR 
clouds.

We have used the line ratios of [Ne\,VI]/[O\,IV] and [Ne\,VI]/[Ne\,II]
as well as the strength of the PAH features as independent indicators for the
identification of composite sources, i.e. sources in which
star-forming regions within the aperture contribute to the fluxes of low lying
fine structure lines. 
As a new tool for determining the properties of dusty galaxies
we provide mid-IR line ratio diagrams, like [O\,IV]25.9$\mu$m/[Ne\,II]12.8$\mu$m,  
which can be used
to distinguish between emission excited by active nuclei and emission from
(circum-nuclear) star forming regions. 
In addition, we present two-dimensional diagnostic diagrams that are fully analogous to 
classical optical diagnostic diagrams, but better suited for objects with high extinction.
Several combinations of lines like [O\,IV], [S\,IV], [Si\,II], or  [Fe\,II], normalized to 
hydrogen recombination lines, can be used,
allowing to adapt to
redshift range or wavelength coverage of different detectors. 

The high ionization lines like [O\,IV], [Ne\,V], or [Ne\,VI] in Seyfert galaxies
are well correlated with both the mid-IR and far-IR
continuum radiation, the latter correlation showing a somewhat larger scatter than the former. 
Low ionization lines, 
which might be significantly affected
by star formation, show similar correlations.
In particular the [Ne\,II]-to-FIR correlation is very similar in both Seyfert and 
starburst galaxies. On the one hand,
this makes it difficult to quantify the dominant contributor to the 
bolometric luminosities of Seyfert host galaxies directly from such ratios of
MIR lines to FIR continuum.
On the other hand, it makes [Ne\,II] a valuable reference line for comparisons
of line ratios between different sources. We have used this to construct simple
linear mixing models of AGN contribution to the bolometric luminosities of composite sources
for the ratios [O\,IV]/[Ne\,II] and [Ne\,V]/[Ne\,II]. 

\begin{acknowledgements}
We wish to thank
Julia Morfill for her help with the data reduction.
SWS and the ISO Spectometer Data Center at MPE are supported by
DLR under grant 50 QI 0202 and by Verbundforschung under grant 50 OR 9913 7. 
This research was supported in part by the German-Israeli Foundation
(grant I-0551-186.07/97).
The ISO Spectral Analysis Package 
(ISAP) is a joint development by the LWS and SWS Instrument Teams and Data 
Centers. Contributing institutes are CESR,
IAS, IPAC, MPE, RAL and SRON.
This research has made use of the NASA/IPAC Extragalactic Database
(NED) which is operated by the Jet Propulsion Laboratory, California Institute 
of Technology, under contract with the National Aeronautics and Space 
Administration. 
\end{acknowledgements}

%
%
%
%
%
%
%


\begin{thebibliography}{}

\bibitem[1999]{AlexanderD} Alexander, D.~M., Efstathiou, A., Hough, J.~H.~et al.\ 1999a, 
MNRAS, 310, 78
et al. \ 1999b, MNRAS, 303, L17
\bibitem[Alexander et al.(1999)]{1999ApJ...512..204A} Alexander, T., Sturm, E., Lutz, D. 
et al.\ 1999, ApJ, 512, 204
\bibitem[Alexander et al.(2000)]{2000ApJ...536..710A} Alexander, T., Lutz, D., Sturm, E.
et al.\ 2000, ApJ, 536, 710
\bibitem[1998]{axon98} Axon D., Marconi A., Capetti A.~et al. 
1998, ApJ, 496, L75
\bibitem[1981]{BPT} 
  Baldwin, J.~A., Phillips, M.~M., \& Terlevich, R.\ 1981, PASP, 93, 5  
\bibitem[2000]{berrington00} Berrington K.A., Nakazaki S., \& Norrington P.H.
2000, A\&AS, 142, 313
\bibitem[1996]{binette96} Binette L., Wilson A.S., \& Storchi-Bergman T.
  1996, A\&A, 312, 365
\bibitem[1997]{binette97} Binette L., Wilson A.S., Raga A., 
  \& Storchi-Bergmann T. 1997, A\&A, 327, 909
\bibitem[1993]{cameron93} Cameron M., Storey, J.~W.~V., Rotaciuc, V., et al. 1993, ApJ, 419, 136 
\bibitem [Chiar et al.(2000)]{2000ESASP.456...45C} Chiar, J.~E., Tielens, A.~G.~G.~M., Whittet, D.~C.~B., 
et al.\ 2000, ISO beyond the peaks: The 2nd ISO workshop on analytical spectroscopy.~Eds.~A.~Salama, M.F.Kessler  K.~Leech \& B.~Schulz.~ESA-SP 456., 456, 45
\bibitem[Cid Fernandes et al.(2001)]{2001ApJ...558...81C} Cid Fernandes, R., Heckman, T., Schmitt, H., 
Delgado, R.~M.~G., \& Storchi-Bergmann, T.\ 2001, ApJ, 558, 81
\bibitem[Clavel et al.(2000)]{2000A&A...357..839C} Clavel, J., Schulz, B., Altieri, B.~et al.
  \ 2000, A\&A, 357, 839
\bibitem[Contini \& Viegas(2001)]{2001ApJS..137...75C} Contini, M.~\& Viegas, S.~M.\ 2001, ApJS, 137, 75 
\bibitem[de Graauw et al. (1996)]{1996A&A...315L..49D} De Graauw, T., Haser, L.~N., Beintema, D.~A., et al. 
  1996, A\&A, 315, L49 
\bibitem[de Grijp, Miley, Lub, \& de Jong(1985)]{1985Natur.314..240D} de 
Grijp, M.~H.~K., Miley, G.~K., Lub, J., \& de Jong, T.\ 1985, Nature, 314, 240 
\bibitem[de Grijp, Lub, \& Miley(1987)]{1987A&AS...70...95D} de Grijp, 
M.~H.~K., Lub, J., \& Miley, G.~K.\ 1987, A\&AS, 70, 95 
\bibitem[Desert, Boulanger, \& Puget(1990)]{1990A&A...237..215D} Desert, 
F.-X., Boulanger, F., \& Puget, J.~L.\ 1990, A\&A, 237, 215 
 A\&AS, 123, 159
\bibitem[Genzel et al.(1998)]{1998ApJ...498..579G} Genzel, R., Lutz, D., Sturm, E., et al.\ 1998, ApJ, 498, 579
\bibitem[GenzelCesarsky (2000)]{GenzelCesarsky} Genzel, R.~\&  Cesarsky, C.~J.\ 2000, ARA\&A, 38, 761 
\bibitem[Giveon et al.(2002)]{2002ApJ...566..880G} Giveon, U., Sternberg, 
A., Lutz, D., Feuchtgruber, H., \& Pauldrach, A.~W.~A.\ 2002, ApJ, 566, 880 
\bibitem[Gonz{\' a}lez Delgado, Heckman, \& Leitherer(2001)]{2001ApJ...546..845G} Gonz{\' a}lez Delgado, 
R.~M., Heckman, T., \& Leitherer, C.\ 2001, ApJ, 546, 845
\bibitem[Heckman et al.(1997)]{1997ApJ...482..114H} Heckman, T.~M., Gonzalez-Delgado, R., 
  Leitherer, C.~et al. 
  \ 1997, ApJ, 482, 114
\bibitem[Kessler et al. (1996)]{1996A&A...315L..27K} Kessler, M. F., Steinz, J.~A., Anderegg, M.~E., et al. 
  1996, A\&A, 315, L27 
\bibitem[Kraemer, Turner, Crenshaw, \& George(1999)]{1999ApJ...519...69K} Kraemer, S.~B., Turner, T.~J.,   Crenshaw, D.~M., \& George, I.~M.\ 1999, ApJ, 519, 69
\bibitem[Kraemer, Crenshaw, \& Gabel(2001)]{2001ApJ...557...30K} Kraemer, S.~B., Crenshaw, D.~M., \& Gabel,   J.~R.\ 2001, ApJ, 557, 30
\bibitem[Kriss et al.(1992)]{1992ApJ...392..485K} Kriss, G.~A., Davidsen, A.~F., 
  Blair, W.~P.~et al.\ 1992, ApJ, 392, 485
\bibitem[Lahuis et al.(1998)]{1998adass...7..224L} Lahuis, F., Wieprecht, E., Bauer, O.~H.~et al.\ 1998, 
  ASP Conf.~Ser.~145: Astronomical Data Analysis Software and Systems VII, 7, 224
\bibitem[1994]{lennon94} Lennon D.J.~\& Burke V.M. 1994, A\&AS, 103, 273
\bibitem[Lutz, Kunze, Spoon, \& Thornley(1998)]{1998A&A...333L..75L} Lutz, D., Kunze, D., Spoon, H.~W.~W., \&   Thornley, M.~D.\ 1998, A\&A, 333, L75
\bibitem[Lutz et al.(2000)a]{2000ApJ...530..733L} Lutz, D., Genzel, R., Sturm, E.~et al.\ 2000, 
  ApJ, 530, 733
\bibitem[Lutz et al.(2000)b]{2000ApJ...536..697L} Lutz, D., Sturm, E., Genzel, R.
  et al. \ 2000, ApJ, 536, 697
\bibitem[1996]{marconi96} Marconi A., van der Werf P.P., Moorwood A.F.M.,
\& Oliva E. 1996, A\&A, 315, 335
\bibitem[Moorwood et al.(1996)]{1996A&A...315L.109M} Moorwood, A.~F.~M., Lutz, D., Oliva, E.
  et al.\ 1996, A\&A, 315, L109
\bibitem[1979]{nussbaumer79} Nussbaumer H.~\& Rusca C. 1979, A\&A, 72, 129
\bibitem[Oliva, Salvati, Moorwood, \& Marconi(1994)]{1994A&A...288..457O} Oliva, E., Salvati, M., 
  Moorwood, A.~F.~M., \& Marconi, A.\ 1994, A\&A, 288, 457
\bibitem[Oliva et al.(1999)]{1999A&A...343..943O} Oliva, E., Moorwood, A.~F.~M., Drapatz, S., 
  Lutz, D., \& Sturm, E.\ 1999, A\&A, 343, 943
\bibitem[Prieto \& Viegas(2000)]{2000ApJ...532..238P} Prieto, M.~A.~\& Viegas, S.~M.\ 2000, ApJ, 532, 238
\bibitem[Prieto et al.(2002)]{astro-ph/0109083} Prieto, M.~A., Perez Garcia, A.~M., 
  \& Rodriguez Espinosa, J.~M.\ 2002, astro-ph/0109083
\bibitem[Rigopoulou et al.(1999)]{1999AJ....118.2625R} Rigopoulou, D., Spoon, H.~W.~W., Genzel,  
  R.~et al. \ 1999, AJ, 118, 2625
\bibitem[Rigopoulou et al.(2002)]{2002A&A} Rigopoulou, D., Kunze, D., Lutz, D.,
  Genzel, R., \& Moorwood, A.~F.~M. \ 2002, astro-ph/0206135
\bibitem[2001]{salas01} Salas J.B., Pottasch S.R., Beintema D.A., 
\& Wesselius P.R. 2001, A\&A, 367, 949
\bibitem[Sanders \& Mirabel(1996)]{1996ARA&A..34..749S} Sanders, D.~B.~\& Mirabel, I.~F.\ 1996, 
  ARA\&A, 34, 749
\bibitem[Siebenmorgen, Moorwood, Freudling, \& Kaeufl(1997)]{1997A&A...325..450S} Siebenmorgen, 
R., Moorwood, A.,  Freudling, W., \& Kaeufl, H.~U.\ 1997, A\&A, 325, 450
\bibitem[Spinoglio \& Malkan(1992)]{1992ApJ...399..504S} Spinoglio, L.~\& Malkan, M.~A.\ 1992, 
ApJ, 399, 504
\bibitem[Spinoglio et al. (2000)]{Spinoglio} Spinoglio, L., Benedettini, M., De Troia, G.~et al.\ 
  2000, ISO beyond the peaks: The 2nd ISO workshop on analytical 
  spectroscopy.~Eds.~A.~Salama, M.F.Kessler, K.~Leech \& B.~Schulz.~ESA-SP 
  456., 456, 261 
\bibitem[Spoon et al. (2002)]{Spoon} Spoon, H.W.W., Keane, J.V., Tielens, A.G.G.M.~et al., 2002, 
  astro-ph/0202163 
\bibitem[Sturm et al.(1996)]{1996A&A...315L.133S} Sturm, E., Lutz, D., Genzel, R.~et al.\ 1996, 
  A\&A, 315, L133
\bibitem[Sturm et al.(1998)]{1998adass...7..161S} Sturm, E., Bauer, O.~H., Brauher, J.~et al.\ 1998, 
  ASP Conf.~Ser.~145: Astronomical Data Analysis Software and Systems VII, 7, 161
\bibitem[Sturm et al.(1999)]{1999ApJ...512..197S} Sturm, E., Alexander, T., Lutz, 
  D.~et al.\ 1999, ApJ, 512, 197
\bibitem[Tran et al.(2001)]{2001ApJ...552..527T} Tran, Q.~D., Lutz, D., Genzel, R., et 
  al.\ 2001, ApJ, 552, 527 
\bibitem[1987]{VO} Veilleux, S.~\& 
  Osterbrock, D.~E.\ 1987, ApJS, 63, 295 
\bibitem[Veilleux(1991)]{1991ApJS...75..357V} Veilleux, S.\ 1991, ApJS, 75, 357
\bibitem[Veilleux et al.(1995)]{1995ApJS...98..171V} Veilleux, S., Kim, D.-C., Sanders, D.~B., Mazzarella,   J.~M.,  \& Soifer, B.~T.\ 1995, ApJS, 98, 171
\bibitem[Veilleux, Kim, \& Sanders(1999)]{1999ApJ...522..113V} Veilleux, S., Kim, D.-C., \& Sanders, D.~B.\   1999, ApJ, 522, 113
\bibitem[Voit(1992)]{1992ApJ...399..495V} Voit, G.~M.\ 1992, ApJ, 399, 495
\bibitem[Vrtilek \& Carleton(1985)]{1985ApJ...294..106V} Vrtilek, J.~M.~\& Carleton, N.~P.\ 1985, ApJ, 294, 106. 
\bibitem[Whittle(1985)]{1985MNRAS.213....1W} Whittle, M.\ 1985, MNRAS, 213, 1
\bibitem[Wieprecht et al.(1998)]{1998adass...7..279W} Wieprecht, E., Lahuis, F., Bauer, O.~H.~et al.\ 1998, 
  ASP Conf.~Ser.~145: Astronomical Data Analysis Software and Systems VII, 7, 279


\end{thebibliography}
\end{document}